\documentclass{aa}
\usepackage[varg]{txfonts}
\usepackage{natbib}
\bibpunct{(}{)}{;}{a}{}{,} % to follow the A&A style
\usepackage{graphicx,setspace,tabularx,stmaryrd,amsmath,subcaption}
\begin{document}

\title{Chemistry in Disks.}

\subtitle{IX. Observations and modeling of HCO$^+$ and DCO$^+$ in DM Tau.\thanks{Based on observations carried out with the IRAM Plateau de Bure Interferometer. IRAM is supported by INSU/CNRS (France), MPG (Germany) and IGN 
(Spain).}}

\author{R. Teague\inst{1}
  \and D. Semenov\inst{1}
  \and S. Guilloteau\inst{2,3}
  \and Th. Henning\inst{1}
  \and A. Dutrey\inst{2,3}
  \and V. Wakelam\inst{2,3}
  \and E. Chapillon\inst{2,3,4}
  \and V. Pietu\inst{4}}

\institute{Max-Planck-Institut f\"{u}r Astronomie, K\"{o}nigstuhl 17, D-69117 Heidelberg, Germany\\ 
email: teague@mpia.de
\and Univ. Bordeaux, LAB, UMR 5804, F-33270, Floirac, France
\and CNRS, LAB, UMR 5804, F-33270 Floirac, France
\and IRAM, 300 Rue de la Piscine, F-38046 Saint Martin 
d'H\`{e}res, France}

\date{Received 4 November 2014 / Accepted 29 December 2014}

\abstract 
  %Context
  {}
  %Aims
  {We study the deuteration and ionization structure of the DM~Tau disk via interferometric 
observations and modeling of the key molecular ions, HCO$^+$ and DCO$^+$.}
  %Methods
  {The Plateau de Bure Array is used to observe DM~Tau in lines of HCO$^+$ (1-0), (3-2) and DCO$^+$ 
(3-2) with a $\sim 1.5^{``}$ angular and $\sim 0.2$\,km~s$^{-1}$ spectral resolution. Using a 
power-law fitting approach the observed column densities profiles are derived and thus the isotopic ratio
$R_{\rm D} =$ DCO$^+$/\,HCO$^+$. Chemical modeling allowed an exploration of the sensitivity of 
HCO$^+$ and DCO$^+$ abundances to physical parameters out with temperature. A steady state 
approximation was employed to observationally constrain the ionization fraction $x(e^-)$.}
  %Results
{Fitting of radiative transfer models suggests that there is a chemical hole 
in HCO$^+$ and DCO$^+$, extending up to 50~AU from the star. More work is required to discern the 
cause of this. The observed column densities of HCO$^+$ and DCO$^+$ at 100~AU were 
$\left(9.8^{+0.3}_{-0.7}\right)\times10^{12}$ and $(1.2 \pm 0.7)\times10^{12}$\,cm$^{-2}$ 
respectively. Where both HCO$^+$ and DCO$^+$ were present, $R_{\rm D}$ was found to increase 
radially from 0.1 at 50~AU to 0.2 at 450~AU. This behaviour was well reproduced by the chemical 
model. The X-ray luminosity of the central star, the interstellar UV and CO depletion were found to 
be the most important physical parameters controlling the abundances of HCO$^+$ and DCO$^+$.  
Differences in the vertical extent of HCO$^+$ and DCO$^+$ molecular layers resulted in different 
responses to changing physical parameters, manifesting as radial gradients in $R_{\rm D}$. The 
ionization fraction was found to be $x(e^-) \sim 10^{-7}$ in the molecular layer, comparable to the 
disk averaged value. Modeling shows that while HCO$^+$ is the most dominant charged molecular ion in 
our disk model, atomic ions, such as C$^+$, S$^+$, H$^+$, Na$^+$ and Mg$^+$, dominate the charge in both the 
molecular layer and disk atmosphere.}
  %Conclusion
  {A high value of $R_{\rm D}$ is indicative of continued deuterium fractionation in a 
protoplanetary disk after pre/protostellar phases. Radial properties of $R_{\rm D}$ can be employed 
to discern the importance of ionization from X-rays and UV, thus necessitating the need for more, 
high resolution observations of DCO$^+$ and other deuterated species in disks. A steady-state 
approach commonly adopted for constraining ionization degree in prestellar cores is not applicable 
for disks where accurate determination of the ionization fraction in the molecular layer requires 
knowledge of the atomic ions present as molecular ions are relatively sparse.}


\titlerunning{Chemistry in Disks IX}
\authorrunning{Teague et al.}

\maketitle

%===================================================================================================
\section{Introduction}

In the view of recent exciting discoveries of various extrasolar planets with the Kepler satellite, 
including Earth-like planets in a habitable zone \citep{2014Sci...344..277Q}, and recent 
ground-breaking Subaru, Herschel, and ALMA observations of protoplanetary 
disks \citep[e.g.,][]{Oeberg_ea10a,2013A&A...553A...5S,2011ARA&A..49...67W,fedele2013,
2013ApJ...762...48G,2013Sci...340.1199V,2014ApJ...782...62R}, it is with great anticipation that 
we begin to unravel the planet formation process. Interferometric observations of protoplanetary 
disks (PPDs) provide a crucial tool in the quest to understand these enigmatic objects. 

The high-resolution measurements of the dust continuum and molecular emission lines of various 
optical thicknesses allow for probes of physics and chemistry in distinct disk regions \citep[for 
recent reviews, see][]{2013ChRv..113.9016H,2014Sci...344..277Q}. To make sense of these data one 
has to employ disk physical/thermo-chemical  models that predict the thermal and density structures 
of disks \citep[e.g.,][]{Woitke_ea09,2013ApJ...766....8A},  gas-grain chemical models 
 to calculate molecular distributions 
\citep[e.g.,][]{1998A&A...338..995W,ah1999,Semenov_ea10,Walsh_ea10}, and radiative transfer 
modeling to simulate molecular lines \citep[e.g. ARTIST\footnote{\url{http://youngstars.nbi.dk/artist/Welcome.html}}, RADMC-3D\footnote{\url{http://www.ita.uni-heidelberg.de/~dullemond/software/radmc-3d/}},][]{RATRAN,2002A&A...395..373V,Pinte_ea06,Pavlyuchenkov_ea07a,2014ascl.soft02014J}.
Another approach is to perform an iterative fitting of 
the observed spectra/interferometric visibilities, applying simpler power-law models of disk 
physical and molecular structure coupled to a fast  LTE/LVG radiative transfer model 
\citep[e.g.,][]{GD98,Pietu_ea07,2008ApJ...681.1396Q,Rosenfeld+etal_2013}. A combination of these 
approaches have been used in series of articles by our ``Chemistry In Disks'' (CID) consortium 
\citep{2007A&A...464..615D,Schreyer_ea08,Henning_ea10,Semenov_ea10,Dutrey_ea11a,Guilloteau_ea12a}.  

In this paper we present interferometric observations of HCO$^+$ J = (3-2), J = (1-0) and DCO$^+$ J 
= (3-2) in DM~Tau, a bona fide T-Tauri star in the Taurus-Auriga star-forming region. Owing to its 
large size, $\sim 800$~AU, and moderate inclination, $i \approx 35^{\degr}$, it has become one of 
the best studied PPDs. In one of the first millimeter studies of PPDs, 
\citet{1994A&A...291L..23G} used the $^{12}$CO and $^{13}$CO J = (2-1) lines to derive a disk 
mass of $1.4 \times 10^{-3} \mathrm{M}_{\sun}$ and disk radius of $\approx 700$\,au. Later, high 
spectral- and spatial-resolution observations of $^{12}$CO, $^{13}$CO and C$^{18}$O have allowed the 
detection of a vertical temperature gradient within the disk and the presence of very cold, $\sim 
10-15$~K CO  gas \citep{2003A&A...399..773D}, a better constrained disk mass of 
0.05~$\mathrm{M}_{\sun}$ and an outer radius in CO of $\approx 800$~AU.

\citet{Pietu_ea07} used higher resolution observations of these lines supplemented with HCO$^+$ J = 
(1-0) to better constrain the disk physical structure and  kinematics in the radial direction.  They 
confirmed the presence of the vertical gas temperature gradient in the DM~Tau system, found that CO 
has an extended distribution in vertical direction, and that the slope of the CO surface 
distribution changes its value with radius.

In the first CID paper of \citet{2007A&A...464..615D}, a sensitive observation of N$_2$H$^+$ and 
HCO$^+$ towards three disks (including DM~Tau) was performed with PdBI, followed by advanced 
physico-chemical modeling. It was found that HCO$^+$ is a major polyatomic ion in disks, and that 
its column density agrees with the modeled values at an evolutionary stage of a few million years. 
The ionization degree in the HCO$^+$ molecular layer was also derived, $\sim 2\times 10^{-9}$.

In the next CID paper by \citet{Schreyer_ea08} the chemical content of the DM~Tau disk was compared 
to the disk around a hotter Herbig A0 star: AB~Aur. We found that while the AB~Aur disk possesses 
more CO, it is less abundant in other, more complex molecular species compared to the DM Tau disk. 
This finding gives a hint that high-energy radiation from the central star may be important not 
only for disk thermal structure but also for its chemical complexity \citep[see 
also][]{2011ApJ...732..106F,Fedele_ea13a,2011ApJ...734...98O}.

Another way to better characterize the thermal structure of PPDs is to observe and analyze 
deuterated species. Unfortunately, the key species for deuterium chemistry that can be detected at 
submillimeter wavelengths, namely ortho-H$_2$D$^+$ and para-D$_2$H$^+$, have been only observed in 
cold prestellar cores 
\citep{2006A&A...454L..55H,2006A&A...454L..59H,2011A&A...526A..31P,2006ApJ...645.1198V,
2011A&A...526A..31P} and have not yet been firmly detected in disks 
\citep{2007A&A...471..187A,Chapillon_ea11}. Thus, to fully characterize deuterium chemistry in 
disks other more readily observable tracers, such as DCO$^+$, DCN, DNC, N$_2$D$^+$, along with 
their major isotopologues, must be used.

%DM Tau Stellar and Disk Properties Table
\begin{table}
\caption{Stellar and disk properties of DM~Tau with values taken from 
\citet{2007A&A...464..615D} and \citet{Henning_ea10}.}
\label{tab:Dm_Tau}
\centering
\footnotesize
\onehalfspacing
\begin{tabular}{lc}
\hline \hline
\multicolumn{2}{l}{$\qquad$\textsc{DM~Tau Stellar and Disk Properties}}\qquad\qquad\qquad\\\hline
Right Ascension (J2000)				& 04$^{\mathrm{h}}$33$^{\mathrm{m}}$48$\fs$733 \\
Declination (J2000)				& +18$\degr$10$\arcmin$09$\farcs$89 \\
Spectral Type 					& M1 \\
Effective Temperature (K) 			& 3720 \\
Stellar Luminosity (L$_{\sun}$) 		& 0.25  \\
Accretion Rate (M$_{\sun}$\,yr$^{-1}$)	 	& 2 $\times$ 10$^{-9}$  \\
Disk Mass (M$_{\sun}$)				& 0.05  \\
$R_{\mathrm{out}}$ (au) 			& 800\\
\hline
\end{tabular}

\end{table}

We must use a combination of these molecules as there are two main deuteration pathways possible, 
each with a different range of temperatures where they are most efficient. For example, DCO$^+$ and 
N$_2$D$^+$ fractionation occurs mainly via H$_3^+$ isotopologues at temperatures $\lesssim 
20-30$~K, whereas fractionation of DCN and DNC involves deuterated light hydrocarbon ions such as 
CH$_2$D$^+$ and C$_2$HD$^+$ with a pathway which remains active up to temperatures of $70-80$~K 
\citep[e.g.,][]{1989ApJ...340..906M,AH99b,Albertsson_ea13,2014arXiv1403.7143C}.

The first detection of a deuterated species in a disk was made by \citet{2003A&A...400L...1V}, who 
detected DCO$^+$ in TW~Hya and discerned a disk averaged ratio of $R_{\rm D} \simeq 0.04$, a 
value similar to that found in pre-stellar cores 
\citep[e.g.,][]{2007ARA&A..45..339B,2012A&ARv..20...56C}. \citet{2006A&A...448L...5G} have since 
detected DCO$^+$ in DM~Tau, with a lower ratio of $R_{\rm D} \sim 4 \times 10^{-3}$. A more recent, 
higher angular resolution study of \citet{2008ApJ...681.1396Q} has shown that the DCO$^+$/\,HCO$^+$ 
ratio in the TW~Hya disk increases radially from 0.01 to 0.1 up to a radius of $\approx$ 90 au, 
where it drops off considerably. Later, \citet{2012ApJ...749..162O} have observed isotopologues of 
HCN and HCO$^+$ in TW~Hya with both then SMA and ALMA, finding that the radial distribution of 
DCO$^+$ and DCN is markedly different. While DCN seems to be centrally peaked, DCO$^+$ shows an 
increasing column density with radius. This supports the theoretical predictions that these two 
deuterated ions are synthesized via distinct low- and high-temperature fractionation pathways. 
Recently, \citet{2013A&A...557A.132M} have directly imaged the location of the CO snowline in a 
warmer disk around a Herbig A3 star HD~163296 with ALMA, using the optically thin DCO$^+$ (5-4) line 
as a direct tracer of CO \citep[see also][]{2013Sci...341..630Q}.

Additional information from the analysis of the HCO$^+$ and DCO$^+$ data includes the possibility 
to better constrain the ionization degree than with HCO$^+$ data alone 
\citep[e.g.,][]{2002P&SS...50.1133C}. The ionization degree is a key quantity that enables angular 
momentum transport in disks, thereby regulating their overall evolution and the ability to 
form planets via turbulence. Magnetorotational instability (MRI; \citet{1991ApJ...376..214B}) is 
currently the most widely accepted source of turbulence in disks 
\citep[see, e.g.][]{2012ApJ...761...95F}, although magnetocentrifugal disk winds and other 
non-linear effects such as ambipolar diffusion can also be important. 
\citet{2011ApJ...743..152O} have used the  CO, HCO$^+$, DCO$^+$, and N$_2$H$^+$ lines observed in 
the disk of DM~Tau with SMA and estimated ionization degree through its molecular layer. In the 
region probed by HCO$^+$ ($T \gtrsim 20$~K) the ionization degree was found to be $4\times 
10^{-10}$, whereas in colder, deeper layers where N$_2$H$^+$ and DCO$^+$ abundances are peaked, the 
ionization degree is lower, $\sim 3 \times 10^{-11}$.

The goal of this paper is to build on the previous work of \citet{2011ApJ...743..152O} to better 
understand the thermal and ionization structure of DM~Tau. We will present higher resolution 
observations of HCO$^+$ J = (3-2), J = (1-0) and DCO$^+$ J = (3-2) which, combined with a 
chemical model of DM~Tau, allow us to determine the radial dependance of HCO$^+$ and DCO$^+$ column 
densities. Furthermore, the combination of two rotational lines for HCO$^+$ with different optical 
depths provide us with a tool with which to discern the HCO$^+$ excitation structure of the disk. 

The paper is structured as follows: Sect.~\ref{sec:observational_model} will describe the treatment 
of the observational data and the method in deriving a model, Sect.~\ref{sec:c_model} details the 
computational model used and a description of the resulting `best-fit' model to our observations. 
In Sect.~\ref{sec:discussion} we explore the deuterium fractionation and ionization fraction in the 
disk with a suite of chemical models to aid analysis, before finally summarising our results in 
Sect.~\ref{sec:summary}.

%===================================================================================
\section{Observational Results}
\label{sec:observational_model}

%Channel Maps of Lines
\begin{figure*}
 \begin{flushright}
 \includegraphics[width=0.9763\textwidth]{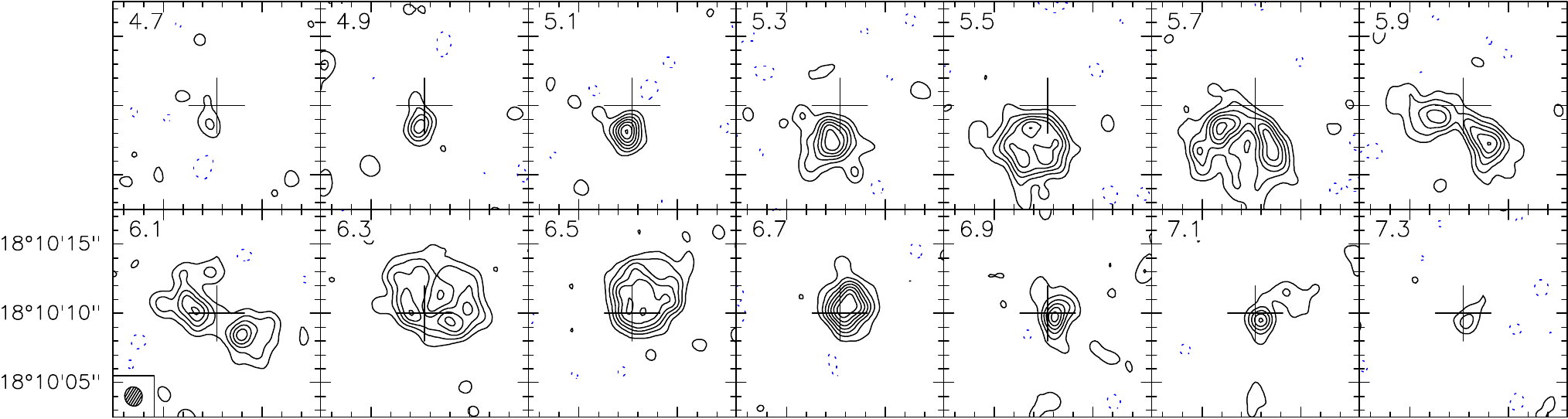}\vspace{1.5mm}
 \includegraphics[width=0.9763\textwidth]{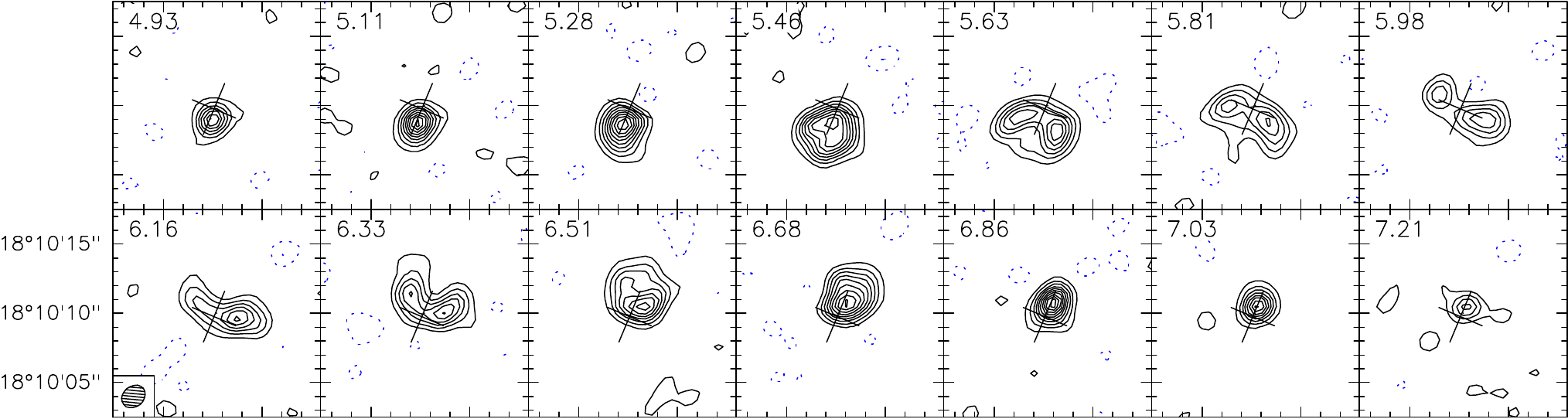}\vspace{1.5mm}
 \includegraphics[width=\textwidth]{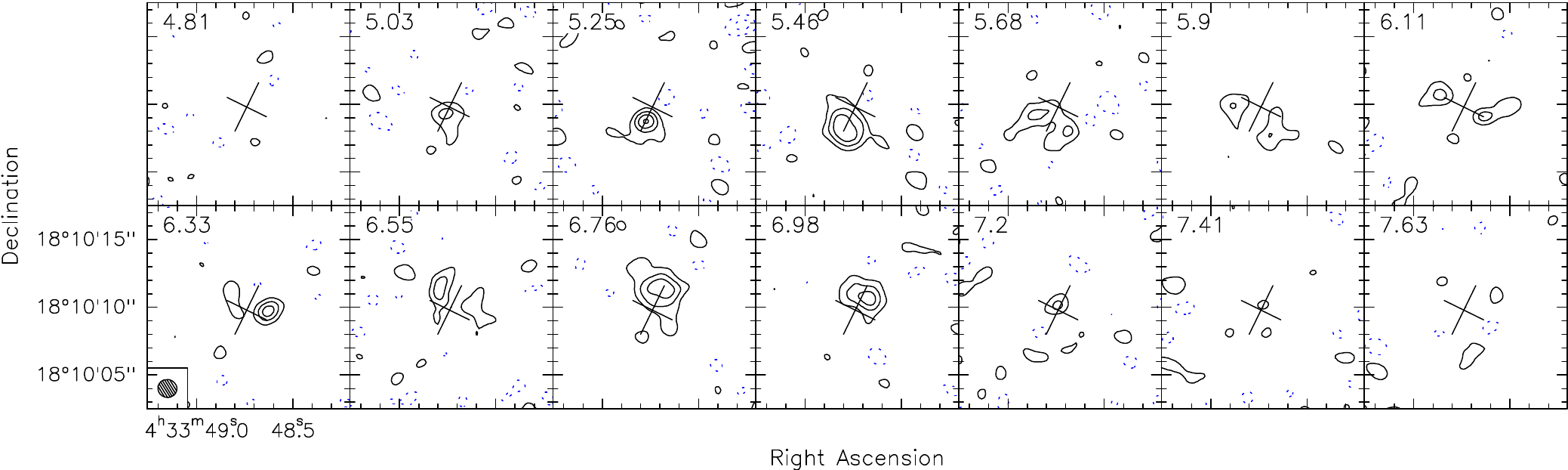}
 \end{flushright}
 \caption{Channel maps of: HCO$^+$ J = (1-0), top, with $\sigma =$ 3.7\,mJy\,/\,beam (0.32~K) and 
contour spacing of 2.5$\sigma$; HCO$^+$ J = (3-2), middle, with $\sigma =$ 100\,mJy\,/\,beam 
(0.64~K) and contour spacing of 2.5$\sigma$; and DCO$^+$ J = 3-2, bottom, with $\sigma =$ 
30\,mJy\,/\,beam (0.42~K) and contour spacing of 2$\sigma$ emission from DM Tau. The beam size, 
shown by the filled ellipse, for each line is 1.4 $\times$ 1.26$\arcsec$, 1.85 $\times$ 
1.49$\arcsec$ and 1.4 $\times$ 1.33$\arcsec$ respectively.  The velocity of each channel shown in 
the top left corner in km\,s$^{-1}$ while the cross in the centre shows the position of the host 
star and the position angle of the major and minor axes of the disk. Dashed contours show negative 
values.} \label{fig:channel_maps}
\end{figure*}

This section describes the process of creating a model of DM~Tau by fitting HCO$^+$ and DCO$^+$ 
line emission.

%===================================================================================

\subsection{PdBI Data}
\label{sec:pdbi_data}

Observations were carried out with the IRAM Plateau de Bure interferometer. 
Table~\ref{tab:Dm_Tau} presents basic stellar properties and disk parameters of DM Tau from 
previous studies. We observed two transitions of HCO$^+$, the J = (1-0) line at 89.18852~GHz and the 
J = (3-2) line at 267.55762~GHz and the J = (3-2) transition of DCO$^+$ at 221.611258~GHz. The 
HCO$^+$ J = (1-0) data included those described in \citet{Pietu_ea07} and were completed by 
longer baselines data (baselines up to 760~m) obtained in late Feb and early March 2008. The DCO$^+$ 
data were obtained between Aug and Dec 2007, with baselines ranging from 15 to 175~m, yielding an 
angular resolution around 1.3$\arcsec$. The HCO$^+$ J = (3-2) data were obtained in Dec 2008, with a 
similar baseline coverage than the DCO$^+$ data.

%===================================================================================

\subsection{Data Reduction}
\label{sec:data_reduction}

We used the IRAM package \texttt{GILDAS}\footnote{http://www.iram.fr/IRAMFR/GILDAS} for data 
reduction and imaging. All data were smoothed to similar spectral resolutions, 0.17 to 
0.20~km~s$^{-1}$ for best comparison. Self-calibration was applied to all three lines and the dust 
thermal continuum was subtracted from the line spectra \citep{2007A&A...464..615D}. 

Channel maps of the emission lines are shown in Fig. \ref{fig:channel_maps}, with contours in levels 
of 2.5$\sigma$ for the HCO$^+$ data and 2$\sigma$ for DCO$^+$. The rms noise values were calculated 
in a line free channel and are 3.7, 100 and 30~mJy\,/\,beam respectively. The need to plot tighter 
2 $\sigma$ contours for DCO$^+$ J = (3-2) highlights the lower intensity of this line relative to 
the HCO$^+$ emission, which is to be expected for a deuterated isotopologue.

%===================================================================================

\subsection{DiskFit Fitting}
\label{sec:diskfit}

%Table of fitted parameters. Two column width.
\begin{table*}
\caption{Best-fit parameters for DM Tau with descriptions of the parameters in the text.}
\footnotesize
\onehalfspacing
\centering
\begin{tabular}{clccccc} \hline \hline
\multicolumn{2}{c}{\textsc{Parameters}}& \textsc{HCO$^+$ J = (1-0)} 	& \textsc{HCO$^+$ J 
= (3-2)} & \textsc{HCO$^+$ Simultaneously}	& \textsc{DCO$^+$ J = (3-2)} & 
\textsc{\footnotesize Continuum} 	\\\hline	
V$_{\mathrm{LSR}}$& (km\,s$^{-1}$)&$6.05\pm0.01$&$6.01\pm0.02$&[6.01]&$6.00\pm0.21$&[6.01]\\ 
$i$&($\degr$)&$34.0\pm2.7$&$33.8\pm0.5$&[34]&$34.5\pm2.1$&[34]\\ 
PA&($\degr$)&$64.31\pm0.57$&$65.7\pm3$&[65]&$65.9\pm1.3$&[65]\\ 
V$_{100}$& (km\,s$^{-1}$)&$2.06\pm0.10$&$2.16 \pm 0.05$&[2.1]&[2.1]&[2.1]\\
$R_{\mathrm{int}}$&(au)&-&$53\pm7$&$49^{+4}_{-3}$&$70\pm 20$&-\\ 
$R_{\mathrm{out}}$&(au)&[750]&$510\pm5$&[800]&$427\pm10$&$173.5\pm0.3$\\ 
d$V$&(km\,s$^{-1}$)&$0.17\pm0.01$&$0.14\pm0.02 $&-&$0.22\pm0.44$&[0.15]\\ 
$h_{100}$&(au)&[16.5]&[16.5]&[16.5]&[16.5]&[16.5]\\
$T_{100}$&(K)&$11.3\pm0.2$&$19.0\pm0.2$&$33.6^{+1.5}_{-1.4}$&[17]&$19.07\pm0.04$\\ 
$q$&&$0.34\pm0.04$&$0.46\pm0.04$&$1.00\pm0.04$&[0.43]&$0.44\pm0.03$\\ 
$\Sigma_{100}$&(cm$^{-2}$)&$(1.9\pm0.8)\times 
10^{14}$ & $\left[2.00\times10^{14}\right]$&$\left(9.8^{+0.3}_{-0.6}\right)\times10^{12}
$&$(1.2\pm0.7)\times 10^{12}$&$(2.2\pm 0.66) \times 10^{23}$\\ 
$p$&&$2.6 \pm 0.8$&[2.5]& $0.82 \pm 0.06$&$0.44 \pm 0.11$&$0.61\pm0.05$\\\hline
\end{tabular}
\singlespacing
\tablefoot{Rotational lines result in parameters for the gas, the continuum probes the dust. Values 
in square brackets were fixed during fitting.}
\label{tab:bestfit_res}
\end{table*}

Following the prescription of \citet{Pietu_ea07}, we assume that the physical properties which 
affect line emission from a disk vary as a radial power law:

\begin{equation}
 a(r) = a_0 \left( \frac{r}{R_0} \right)^{-e_a},
\end{equation}

\noindent where $a_0$ is the parameter value at the reference radius $R_0$. We adopt the standard 
that positive exponents, $e_a$, imply a decrease of the physical quantity with radius. Through this 
prescription line emission in a disk can be described by the following parameters:

\begin{itemize}
 \item $X_0$ ($^{\prime\prime}$) and $Y_0$ ($^{\prime\prime}$), position of the central star;
 \item V$_{\mathrm{LSR}}$ (km\,s$^{-1}$), systemic velocity;
 \item PA ($\degr$) and $i$ ($\degr$), position angle and inclination of the disk;
 \item $V_{v}$ (km\,s$^{-1}$), $R_{v}$ (au), $v$, rotation velocity at reference radius $R_v$, 
typically 100 au, with a power law 
exponent $v$. Perfect Keplerian rotation would yield $v = 0.5$;
 \item $T_m$ (K), $R_T$ (au), $q_m$, gas temperature at the reference radius $R_T$ and respective 
power law exponent;
 \item d$V$ (km\,s$^{-1}$), $e_v$, turbulent component of the line widths and respective power 
law exponent;
 \item $\Sigma_m$ (cm$^{-2}$), $R_{\Sigma}$ (au), $p_m$, gas surface density, reference radius and 
power law exponent;
 \item $R_{\mathrm{in}}$ and $R_{\mathrm{out}}$ (au), inner and outer radius of observed emission;
 \item $h_m$ (au), $R_h$ (au), $e_h$, scale height of the gas, reference radius and its exponent.
\end{itemize}

\noindent Note that the inclination and position angle are chosen in the range $0^{\degr} \leq 
\mathrm{PA} < 360^{\degr}$ and $-90^{\degr} \leq i \leq 90^{\degr}$ such that $V_{100}$ is always 
positive and that position angle refers to the rotation axis. 

It has been shown previously that power laws are reasonably accurate proxies for correct 
descriptions of the large-scale disk properties, such as kinematics (assuming that the self-gravity 
is negligible), kinetic temperature distribution, surface density (assuming the $\alpha$-viscosity 
prescription and a constant accretion rate), and thus, also the disk scale height 
\citep[e.g.,][]{CG97,Pietu_ea07}. 

The corresponding disk parameters were fitted to each of the observed lines with the 
\texttt{DISKFIT} software \citep{GD98}, using a combination of $\chi^2$ minimization and MCMC 
fitting of the observed visibilities in the $uv$-plane. Line emission provided information on the 
gas structure while continuum emission was used to fit the dust structure.

%===================================================================================
\subsection{Observational Results}
\label{sec:observational_results}

The derived best-fit parameters for the three emission lines and the continuum are shown in 
Table~\ref{tab:bestfit_res}. Geometrical properties of the disk (systemic velocity, inclination, 
position angle and Keplerian velocity at 100~AU) were found to be in good agreement with 
previous studies on DM~Tau using the \texttt{DISKFIT} approach 
\citep{GD98,Pietu_ea07,2007A&A...464..615D}.

Due to the uncertainty in optical depth of the HCO$^+$ rotational lines, we analyzed the HCO$^+$ 
images in three different ways. In case A, the J = (1-0) and (3-2) images were fitted independently. 
The derived parameters are presented in Col. 1-2 of Table \ref{tab:bestfit_res}. The HCO$^+$ J = 
(3-2) is largely optically thick, and thus provides a good estimate of the temperature, however, 
the surface density cannot be well constrained from this data. A best fit outer radius of around 
500~AU was found for this line. On the other hand, HCO$^+$ J = (1-0) extends further (to at least 
750~AU), and is mostly optically thin. Thus, the derived temperature heavily relies on the power 
law extrapolation at low radii.  We also note that the J = (3-2) line requires an inner radius 
around 50~AU. A lack of emission from the inner 50~AU may bias the temperature derived
from the J = (1-0) line towards low values.

In case B, we assume the J = (1-0) and (3-2) lines to have the same excitation temperature. We fit 
both transitions simultaneously, and set the outer radii to 750~AU. Results are given in Col. 3. The 
derived temperature law is now much steeper, $q = 1.00 \pm 0.04$, and the surface density law much 
flatter. The derived surface densities are also lower roughly a factor of 2 to 5 between 200 and 
400~AU compared to the separate fits.

Finally, in case C, we relaxed the power law assumption, and fitted the temperature and surface 
densities at 5 different radii, extrapolating by power laws in between. The solution is within the 
errors consistent with case B, and thus we only report this latter case hereafter. 

Neither approach is perfect: in case A, the extrapolation required to derive the temperature profile 
from the J = (1-0) line is hazardous. On the other hand, in case B we neglect the vertical 
temperature gradients which are expected in disks, despite the higher opacity of the J = (3-2) line 
naturally leading to higher excitation temperature compared to the J = (1-0) line. Alternatively, 
if the density is insufficient to thermalize the J = (3-2) transition, we may expect its 
excitation temperature to be lower than that of the J = (1-0). We shall use the results of case B 
hereafter. The very low temperatures derived in the outer part suggest some sub-thermal excitation, 
at least for the J = (3-2) transition beyond 400~AU or so. Note that apparent ``ring-like'' 
distribution of HCO$^+$ J = (3-2) is not due to simple excitation effect, as suggested by 
\citet{2014ApJ...794..123C}. A central hole of 50~AU radius, almost fully devoid of HCO$^+$, is 
required to reproduce the J = (3-2) emission.

The DCO$^+$ line is weaker, and we can only derive the surface density by fixing all other 
parameters guided by those found for HCO$^+$ as it is a reasonable assumption to assume they are 
cospatial in the disk. Note, however, that the derived values are quite insensitive to the assumed 
temperature law. 

To provide a better comparison with the modeled results, we will extrapolate the power laws 
describing column densities and temperatures. However it must be noted that, due to the inner hole 
in HCO$^+$ and smaller outer radius in DCO$^+$ the power laws only provide a good fit to the data in 
the region $50 \la r \la 430$~AU. In order to better comprehend these findings, we perform detailed 
theoretical modeling in the next Section.

%===================================================================================
\section{Computational Model}
\label{sec:c_model}

%Physical structure of disk model.
\begin{figure}\centering
\includegraphics[width=\columnwidth]{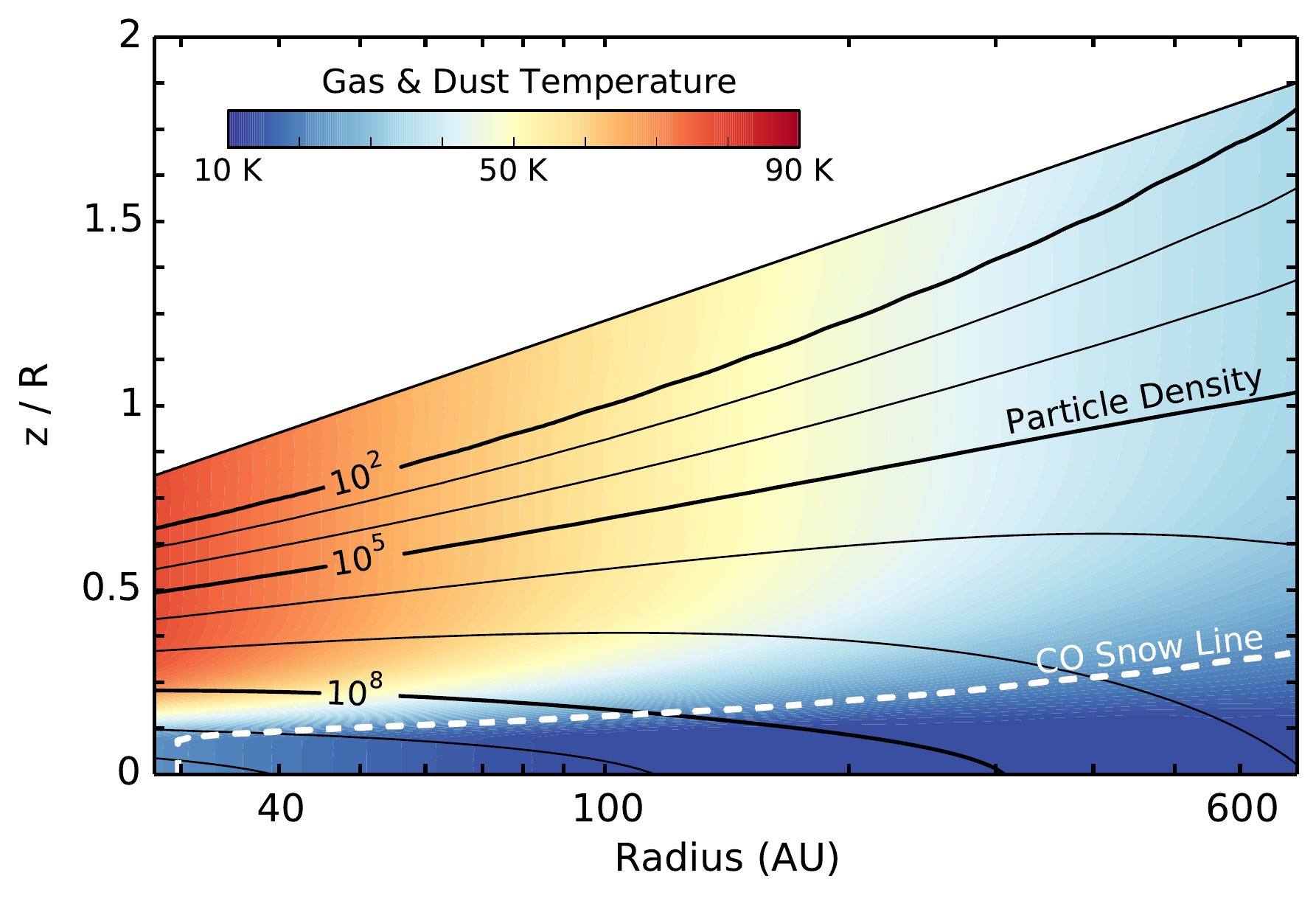}
\caption{Disk physical structure scaled vertically as $z/R$. Colouring shows the coupled gas and 
dust temperature and solid lines show the particle number densities in particles\,cm$^{-3}$. The 
dashed white lines shows the 21\,K isotherm, below which half of CO has frozen out.}
\label{fig:disk_struc}
\end{figure}

%===================================================================================
\subsection{Disk Physical Structure}
\label{sec:disk_phys}

This section describes the methodology of creating a computational chemical disk model of DM~Tau, 
including a description of the physical and chemical parameters used.

The DM~Tau system, at a distance of 140~pc, consists of a single isolated pre-main-sequence 
M0.5-1.5 dwarf ($T_{\rm eff}=3720$~K), with a mass of $0.5-0.65\,\mathrm{M}_\odot$, a radius of 
$1.2\,\mathrm{R}_\odot$, and an accretion rate of $\sim 2-3\times 
10^{-9}\,\mathrm{M}_\sun$~yr$^{-1}$
\citep[][]{Mazzitelli89,Simon_ea00,2014ApJ...780..150M}. It is enshrouded by an extended ($\sim 
800$~AU), cold ($T\ga 10$~K) Keplerian disk \citep{Pietu_ea07}. According to the {\it Spitzer} 
IRS observations \citep{Calvet_ea05}, the inner DM~Tau disk is cleared of small dust ($\la 3-4$~AU) 
and is in a pre-transitional phase. As our interferometric observations have the highest 
sensitivity 
in disk regions  $\ge 30$~AU from the central star, we only consider the chemical evolution outside 
of this radius in our analysis.

The DM~Tau physical disk model is based on a 1+1D steady-state $\alpha$-model similar to that 
of \citet{DAea99}, where equal gas and dust temperatures are assumed. This model was extensively 
used in our previous studies of DM~Tau-like disk chemistry 
\citep[e.g.,][]{Henning_ea10,Semenov_Wiebe11,Albertsson_ea14a}. The disk model has an outer radius  
of $800$~AU, an accretion rate of $2 \times10^{-9}\,\mathrm{M}_\odot$\,yr$^{-1}$, a viscosity 
parameter $\alpha = 0.01$, and a total gas mass of $0.066\,\mathrm{M}_\odot$ 
\citep{2007A&A...464..615D,Henning_ea10,Semenov_Wiebe11}. The dissociating UV radiation of DM~Tau 
is 
represented by the scaled-up interstellar UV radiation field of \citet{G}. The unattenuated stellar 
UV intensity at the radius of 100~AU is $\chi_*(100)=410$ \citep[e.g.,][]{Bergin_ea04}. The X-ray 
luminosity of DM~Tau is taken to be $2 \times 10^{29}$~erg\,s$^{-1}$~\citep[see][]{Semenov_Wiebe11}. 
The calculated disk thermal and density structure is shown in Fig.~\ref{fig:disk_struc}.

%===================================================================================
\subsection{Disk Chemical Model}
\label{sec:disk_chem}

\begin{table*}
\centering                          
\footnotesize
\onehalfspacing
\begin{tabular}{llllllllll}        
\hline\hline
\multicolumn{10}{c}{\textsc{initial abundances for modeling pre-disk evolutionary phase}}\\\hline
ortho-H$_2$&   $0.375$     &He         &   $9.75 (-2)$  &O          &   $1.80 (-4)$  &Na         &   
$2.25 (-9)$  &P          &   $2.16 (-10)$ \\
para-H$_2$ &   $0.125$     &C          &   $7.86 (-5)$  &S          &   $9.14 (-8)$  &Mg         &   
$1.09 (-8)$  &Cl         &   $1.00 (-9)$  \\
HD         &   $1.55 (-5)$ &N          &   $2.47 (-5)$  &Si         &   $9.74 (-9)$  &Fe         &   
$2.74 (-9)$  & \\\hline                              
\end{tabular}

\vspace{0.5cm}

\begin{tabular}{llllllllllll}        
\hline\hline
\multicolumn{12}{c}{\textsc{25 initially most abundant molecules for disk chemical modeling 
including HCO$^+$ and DCO$^+$}}\\\hline
para-H$_2$       &   $3.77 (-01)$  & CO$^*$ &   $4.05 (-05)$  &NH$_3^*$ &   $5.64 (-06)$  &H$^*$     
   &   $6.03 (-07)$  &NO               	&   $2.22 (-07)$  & HCO$^+$       	&   $6.13 (-09)$  \\
ortho-H$_2$      &   $1.23 (-01)$  & CO     &   $3.26 (-05)$  &O        &   $5.59 (-06)$  
&C$_3$H$_2^*$ &   $4.48 (-07)$  &N                	&   $1.36 (-07)$  & DCO$^+$       	&   
$1.25 (-11)$  \\
He               &   $9.75 (-02)$  & O$_2$  &   $1.79 (-05)$  &O$_2^*$  &   $4.12 (-06)$  &OH        
   &   $3.43 (-07)$  &HDO$^*$             	&   $1.35 (-07)$  &               	&   		
  \\
H                &   $5.25 (-04)$  & HD     &   $1.52 (-05)$  &CH$_4^*$ &   $3.64 (-06)$  &H$_2$O    
   &   $2.79 (-07)$  &CO$_2$              	&   $1.32 (-07)$  &               	&   		
  \\
H$_2$O$^*$       &   $5.53 (-05)$  & N$_2$  &   $7.39 (-06)$  &N$_2^*$  &   $1.76 (-06)$  &HNO$^*$   
   &   $2.40 (-07)$  &CO$_2^*$            	&   $1.19 (-07)$  &               	&   		
  \\

\hline                
\end{tabular}

\caption{Atomic and molecular abundances used in the disk chemical model. Top: Initial abundances 
for modeling the pre-disk evolutionary phase Bottom: 25 initially most abundant molecules for disk 
chemical modeling. Note that $a(b)$ should be read as $a \times 10^{b}$ and that $^*$ denote frozen 
species.} 
\label{tab:init_abunds_TMC1}
\end{table*}

The adopted chemical model is based on the advanced \texttt{ALCHEMIC} code 
\citep[see][]{Semenov_ea10} and utilizes the high-temperature, gas-grain deuterium chemistry 
network of \citet{Albertsson_ea13}, with the addition of nuclear spin-state processes for H$_2$, 
H$_2^+$, and H$_3^+$ isotopologues from \citet{2014ApJ...787...44A}. The chemical network without 
deuterated species is based on the osu.2007 ratefile\footnote{See: 
\url{http://www.physics.ohio-state.edu/~eric/research.html}}, with the recent updates to
the reaction rates from Kinetic Database for Astrochemistry (KIDA) \citep{KIDA}. For all 
H-bearing reactions in this network, we derived the corresponding D-bearing reactions following the 
algorithm of \citet[][]{1996MNRAS.280.1046R}. The cloning was not allowed for any species with the 
-OH endgroup.

Primal isotope exchange reactions for H$_3^+$ as well as CH$_3^+$ and C$_{2}$H$_{2}^+$ from 
\citet{2000A&A...361..388R, GHR_02, 2004A&A...424..905R, 2005A&A...438..585R} were included. In 
cases where the position of the deuterium atom in a reactant or in a product was ambiguous, a 
statistical branching approach was used. This deuterium network was further extended by adding 
ortho- and para-forms of H$_2$, H$_2^+$ and H$_3^+$ isotopologues and the related nuclear spin-state 
exchange processes from several experimental and theoretical studies 
\citep{1990JChPh..92.2377G,GHR_02,2004A&A...427..887F,2004A&A...418.1035W,2006A&A...449..621F, 
2009A&A...494..623P, 2009JChPh.130p4302H,2011PhRvL.107b3201H,2013A&A...554A..92S}.

To calculate UV ionization and dissociation rates, the mean FUV intensity at
a given disk location is obtained by summing up the stellar $\chi_*(r)=410 \times (r/100)^{-2}$, where r is in au, and interstellar UV fluxes scaled down by the visual extinction in the radial and 
vertical directions, respectively. Several tens of newer photoreaction rates are adopted from 
\citet{vDea_06}~\footnote{\url{http://www.strw.leidenuniv.nl/~ewine/photo}}. The self-shielding of 
H$_2$ from photodissociation is calculated by  Eq.~(37) from \citet{DB96}. The shielding of CO by 
dust grains, H$_2$, and the CO self-shielding is calculated using a precomputed table of
\citet[][Table~11]{1996A&A...311..690L}.

The stellar X-ray radiation is modeled using observational results of \citet{Glassgold_ea05} and the approximate expressions (7--9) from the 2D Monte Carlo simulations of \citet{zetaxa,1997ApJ...485..920G}. Implementing Eqn. (8) from \citet{1997ApJ...485..920G}, we use an exponent of  $n = 2.81$, a cross section at 1~keV of $\sigma_{-22} = 0.85 \times10^{-22}$~cm$^2$ and total X-ray luminosity of $L_{\rm XR} = 3 \times 10^{29}$~erg~s$^{-1}$, yielding a typical X-ray photon energy of 3~keV. Attenuation of X-rays is calculated from Eqn. (4) in \citet{zetaxa}. The X-ray emitting source is located at 12 stellar radii above the midplane with rates exceeding that of the CRPs in the disk regions above the midplane, particularly, at radii $\sim 100-200$~AU \citep[see also][]{Henning_ea10}

We assume the standard cosmic ray (CR) ionization rate $\zeta_{\rm CR}=1.3\times10^{-17}$~s$^{-1}$ 
and model its attenuation using Eq.~(3) from \citet{2004A&A...417...93S}. Note that we do not 
consider the scattering of low energy CR protons by the heliosphere of DM~Tau, as done in 
\citet{2013ApJ...772....5C}. Ionization due to the decay of short-living radionuclides is taken into 
account, $\zeta_{\rm RN}=6.5\times10^{-19}$~s$^{-1}$ \citep{FG97}, see also 
\citet{2013ApJ...777...28C}.

% Deuterium fraction figures: contour plots and column densities.
\begin{figure*}
 \centering
   \includegraphics[width=1.9\columnwidth]{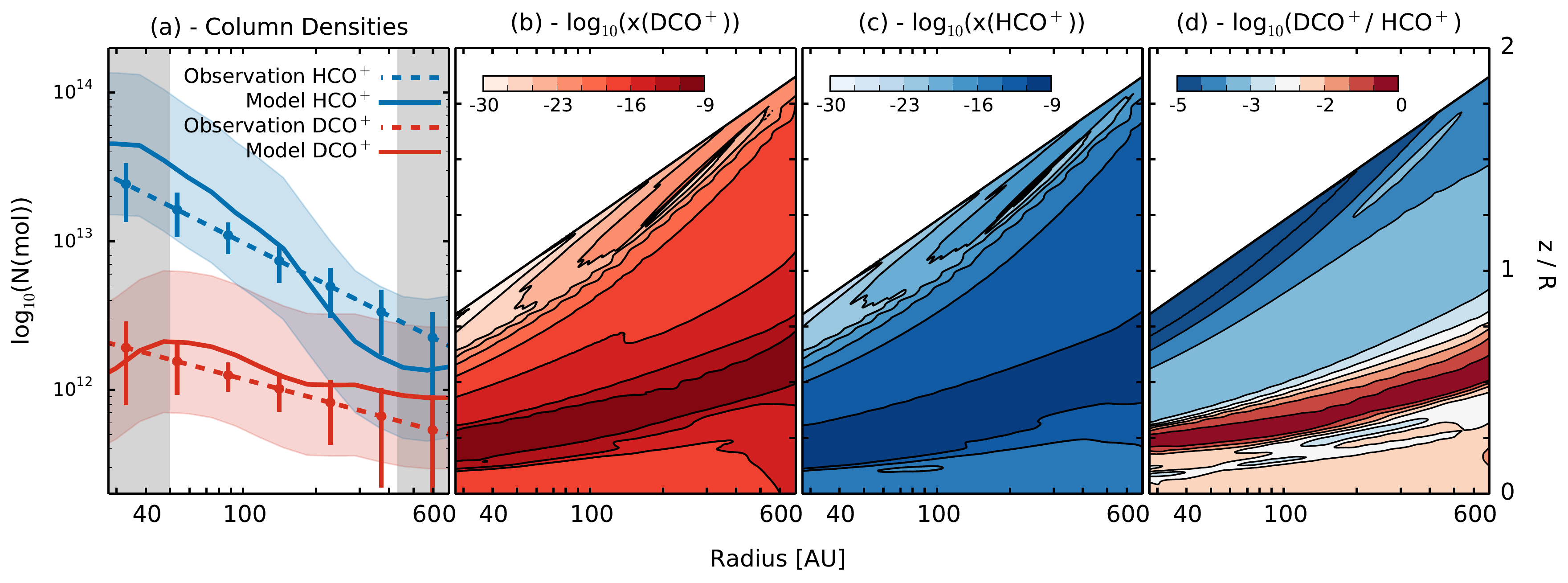}
 \caption{Deuterium fractionation in the disk. (a) Comparisons of HCO$^+$, blue, and DCO$^+$, 
red, from observations, dashed with 3$\sigma$ errors, and the best fit model, solid with associated 
error of a factor of 3. They grey vertical bars show where the observationally derived column densities are 
extrapolated beyond the inner and outer edges found in parameterisation. (b) and (c) show the 
relative abundances of DCO$^+$ and HCO$^+$ in the best fit disk chemical model respectively. (d) 
shows the local DCO$^+$/\,HCO$^+$ ratio in the disk 
model.}
 \label{fig:deuterium_frac}
\end{figure*}

The grain ensemble used to calculate disk physical structure was represented by uniform amorphous 
silicate particles of olivine stoichiometry with density of $3$~g\,cm$^{-3}$ and radius of 
$0.1\,\mu$m. Each grain provides $\approx 1.88\times10^6$ surface sites 
\citep[][]{Bihamea01} for surface recombination that proceeds solely through the classical 
Langmuir-Hinshelwood mechanism \citep[e.g.][]{HHL92}. The gas-grain interactions include sticking of 
neutral species and electrons to dust grains with 100\% probability and desorption of ices by 
thermal, CRP-, and UV-driven processes. We do not allow H$_2$ to stick to grains as it requires 
temperatures of $\la 4$~K. The UV photodesorption yield of $3 \times 10^{-3}$ was adopted 
\citep[e.g.,][]{2009A&A...504..891O,2009A&A...496..281O,Fayolle_ea11a,2013A&A...556A.122F}.
Photodissociation processes of solid species are taken from \citet{Garrod_Herbst06,Semenov_Wiebe11}. 

In addition, dissociative recombination and radiative neutralization of molecular ions on charged 
grains and grain re-charging are taken into account. Upon a surface recombination, we assume there 
is a 1\% probability for the products to leave the grain due to the partial conversion of the 
reaction exothermicity into breaking the surface-adsorbate bond 
\citep{2007A&A...467.1103G,2013ApJ...769...34V}. Following experimental studies on the formation of 
molecular hydrogen on amorphous dust grains by \citet{Katz_ea99}, the standard rate equation 
approach to the surface chemistry was utilized. Overall, the disk chemical network consists of 1268 
species made of 13 elements and 38812 reactions.

The age of the DM~Tau system is poorly constrained, $\sim$ 3--7~Myr \citep[][]{Simon_ea00}. 
In the chemical modeling the age of $5$~Myr was considered. To set initial abundances, we 
calculated chemical evolution in a TMC1-like molecular cloud ($n_{\rm H} = 2\times10^4$~cm$^{-3}$, 
$T = 10$~K, $A_{\rm V}=10$~mag) over 1~Myr. For that, the ``low metals'' elemental abundances of 
\citet{1982ApJS...48..321G,Lea98,2013ChRv..113.8710A} were used, with the equilibrium 3:1 ortho/para 
H$_2$ ratio (hydrogen being fully in molecular form) and deuterium locked in HD molecule (see 
Table~\ref{tab:init_abunds_TMC1}). The resulting abundances of modeling the pre-disk evolutionary 
phase were used as initial abundances for disk chemical modeling as shown in 
Table~\ref{tab:init_abunds_TMC1}.

%===================================================================================================
\subsection{Modelled Results}
\label{sec:bestfit_model}

Figure.~\ref{fig:deuterium_frac}a presents the modeled column densities for HCO$^+$ 
and DCO$^+$ in DM~Tau (dashed lines) and associated 3~$\sigma$ errors with the 
observationally derived values (solid lines) overlain. Both column densities agree well within 
their errors. The grey boxes in Fig~\ref{fig:deuterium_frac} show where the power-law column 
densities are extrapolated beyond the radii where they were observed as to provide a better 
comparison with the chemical model which does not reproduce the inner hole.

Figures~\ref{fig:deuterium_frac}b and c show the relative abundance with respect to H$_2$ of 
DCO$^+$ 
and HCO$^+$ throughout the disk. These clearly demonstrate the stratification of the disk 
with a distinct molecular layer lying $\sim$~0.5 pressure scale heights above the midplane. 
Both molecular layers are relatively co-spatial, in general tracing regions of high gas 
phase CO abundance. They are bounded by the CO snowline towards the midplane, and the 
photodissociation region of CO towards the disk atmosphere. DCO$^+$ occupies a slightly tighter 
vertical range than HCO$^+$ due to it also requiring efficient deuteration. The molecular layer is 
truncated upwards due to the higher gas temperatures, reducing the efficiency of deuterium 
fractionation, while the lower bound is due to reduced HD abundances with which to readily 
transform deuterium into DCO$^+$.

$R_{\mathrm D}$(HCO$^+$) is shown in Fig.~\ref{fig:deuterium_frac}d with local values 
reaching as high as $\sim$~1. Within this region we see pronounced increases of HD and H$_2$D$^+$ 
abundances facilitating a fast transfer of deuterium from HD to DCO$^+$. Despite these locally 
high values, the vertically integrated column density is more sensitive to the denser 
regions closer to the midplane, thereby exhibiting a lower value of $R_{\rm 
D}$(HCO$^+$)~$\sim$~0.1.

Figure~\ref{fig:deuterium_frac_time} shows the modelled values of $R_{\rm D}$ a various time steps 
in the model (blue lines). The observationally derived value, shown by the black dashed line, 
agrees 
qualitatively well with the $t = 5 \times 10^{6}$~yrs value.

%== Discussion =====================================================================
\section{Discussion}
\label{sec:discussion}

In this section we use both the observationally-derived and best-fit modeled column densities, 
complemented with a further suite of chemical models, to explore both deuterium fractionation and 
the ionization fraction in the disk.

%== Deuterium Fractionation ===========================================================

\subsection{Deuterium Fractionation of HCO$^+$}
\label{sec:deuterium_discussion}

% Time dependence of DCO+ / HCO+
\begin{figure}
 \centering
   \includegraphics[width=\columnwidth]{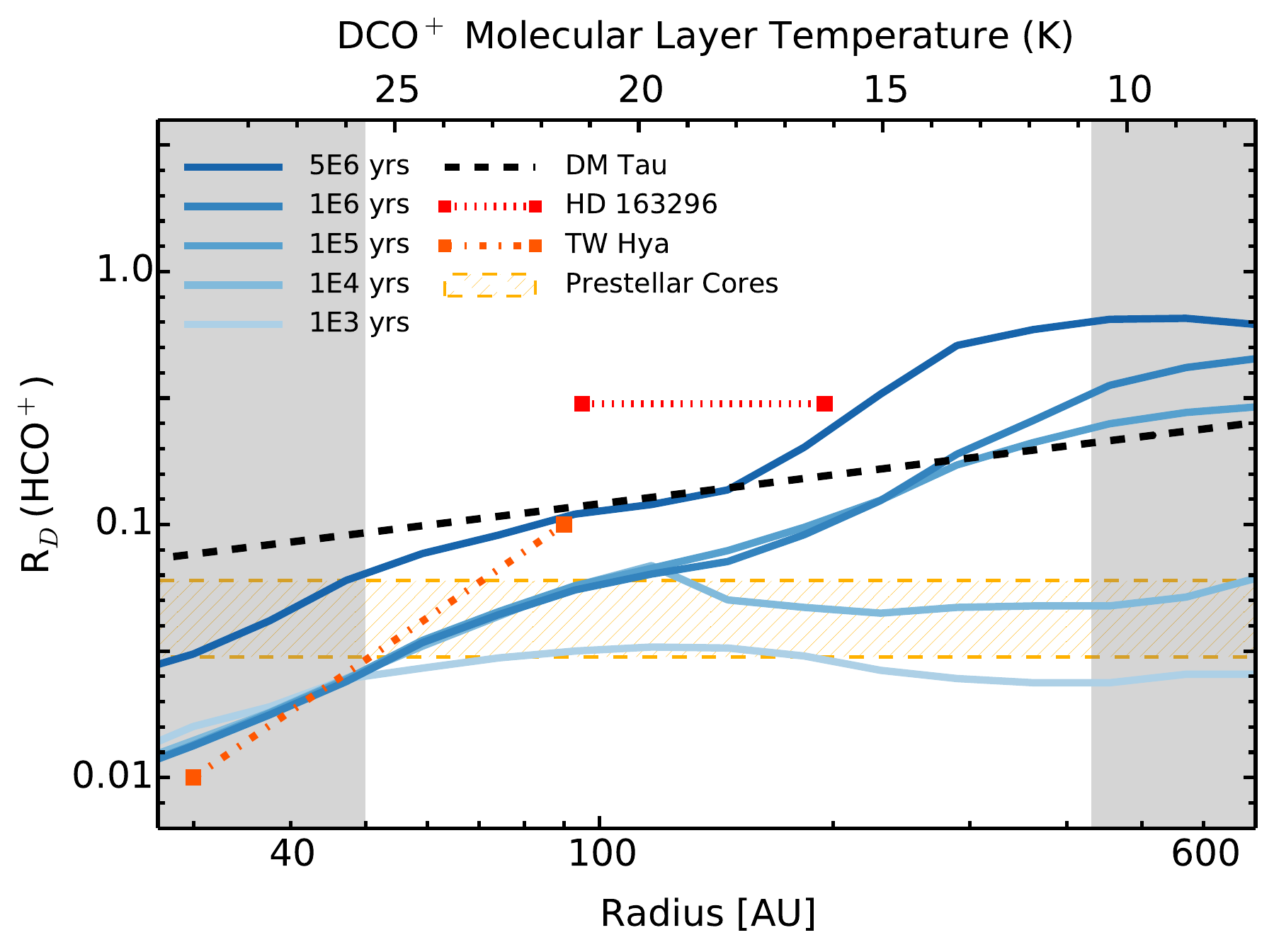}
 \caption{Comparisons of $R_{\rm D}$(HCO$^+$) values. The dashed black line shows the 
observationally derived value in DM Tau and the blue solid lines show the chemical model values at 
different time steps. Typical errors are a factor of 3. They grey vertical bars show where the observationally derived column densities of HCO$^+$ and DCO$^+$ are extrapolated beyond the inner and outer edges found in parameterisation. The orange dash-dotted line shows the value observed in TW~Hya \citep{2008ApJ...681.1396Q}, the red dotted line, $R_{\rm D}$(HCO$^+$) in 
HD~163296 \citep{2013A&A...557A.132M}, and the yellow dashed region, values from a survey of 
prestellar cores 
\citep{1995ApJ...448..207B}. The top y-axis shows the temperature of the DCO$^+$ layer derived 
from our parametric fitting.}
\label{fig:deuterium_frac_time}
\end{figure}

% Radial CO depletion figure and ortho/para evolution.
\begin{figure}
 \centering
 \includegraphics[width=\columnwidth]{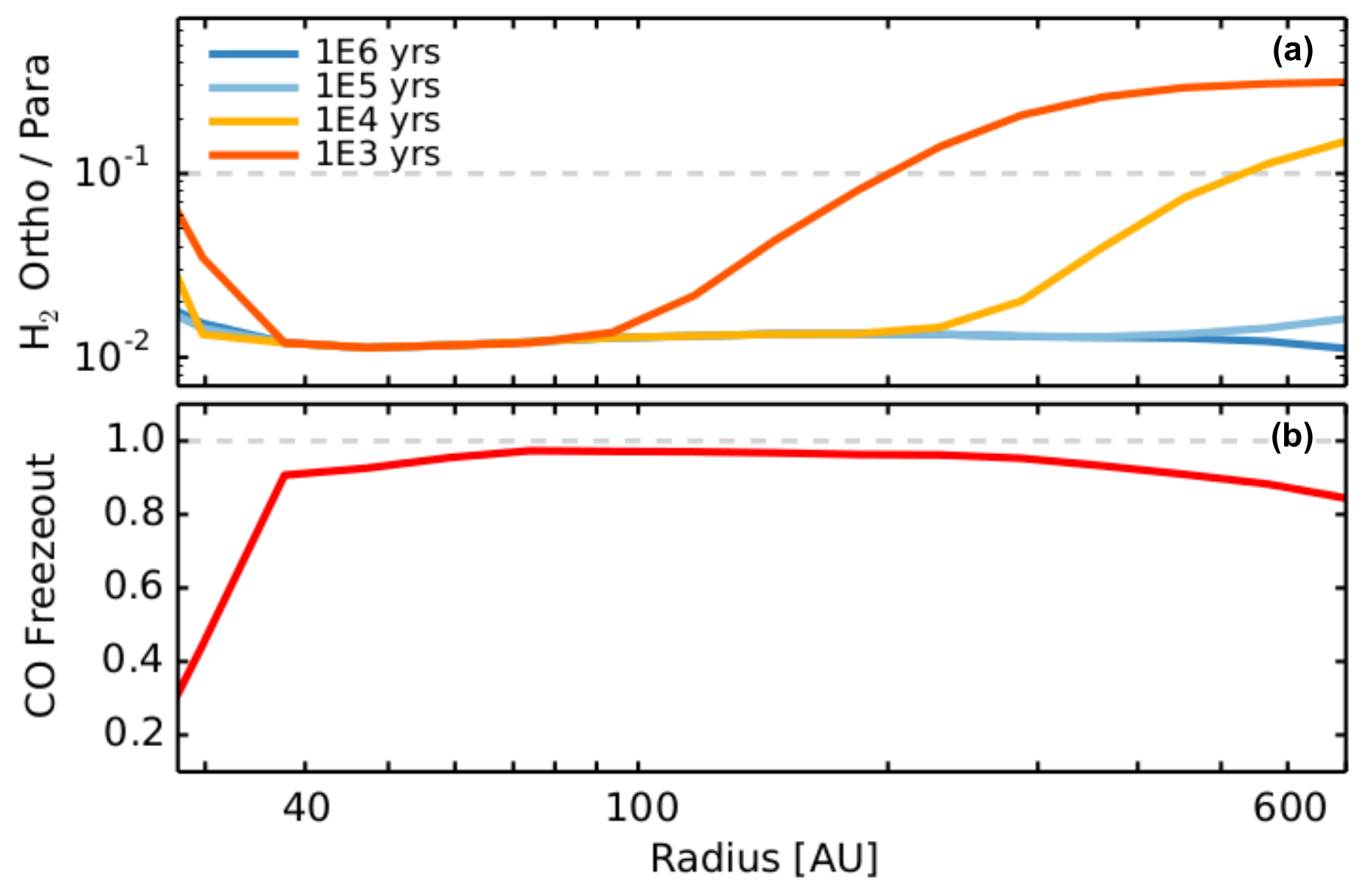}
 \caption{(a) The 
$N$(ortho-H$_2$)\,/\,$N$(para-H$_2$) ratio from the best fit model. After $\sim 10^5$~yrs the ratio 
reaches a steady state. (b) The degree of CO freeze out in our 
model, $N$(CO$_{\mathrm{ice}}$)\,/\,$N$(CO$_{\mathrm{total}}$) 
clearly demonstrating the CO ice line at $r \approx 30$~AU.}\label{fig:co_ice}
\end{figure}

% Calculated molecular abundances w/ observational data.
\begin{figure*}
 \centering
 
   \begin{subfigure}[b]{0.32\textwidth}
   \includegraphics[width=\textwidth]{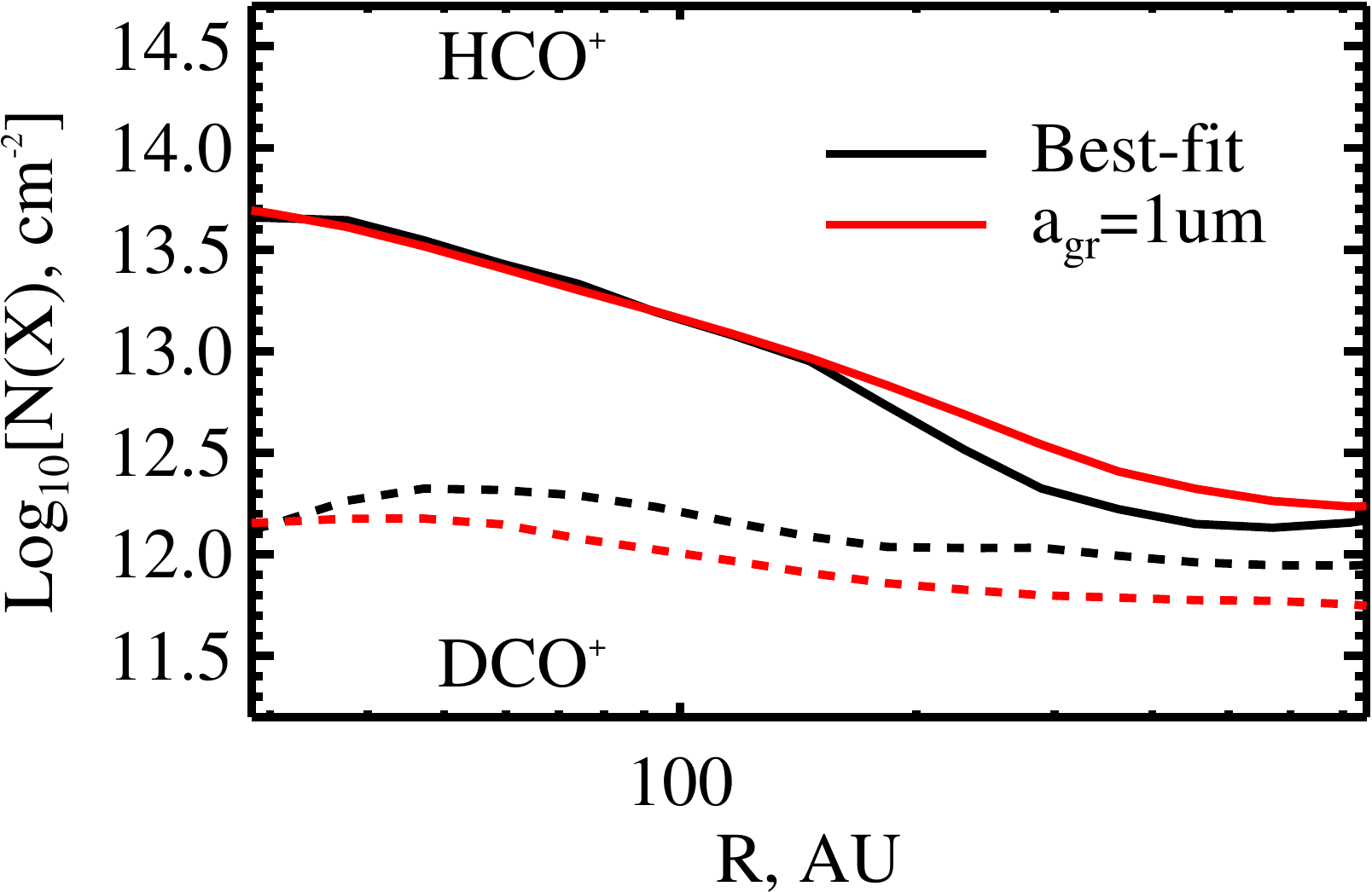}
   \caption{Increased grain size.}
   \label{fig:5grains}
   \end{subfigure}
   ~
   \begin{subfigure}[b]{0.32\textwidth}
   \includegraphics[width=\textwidth]{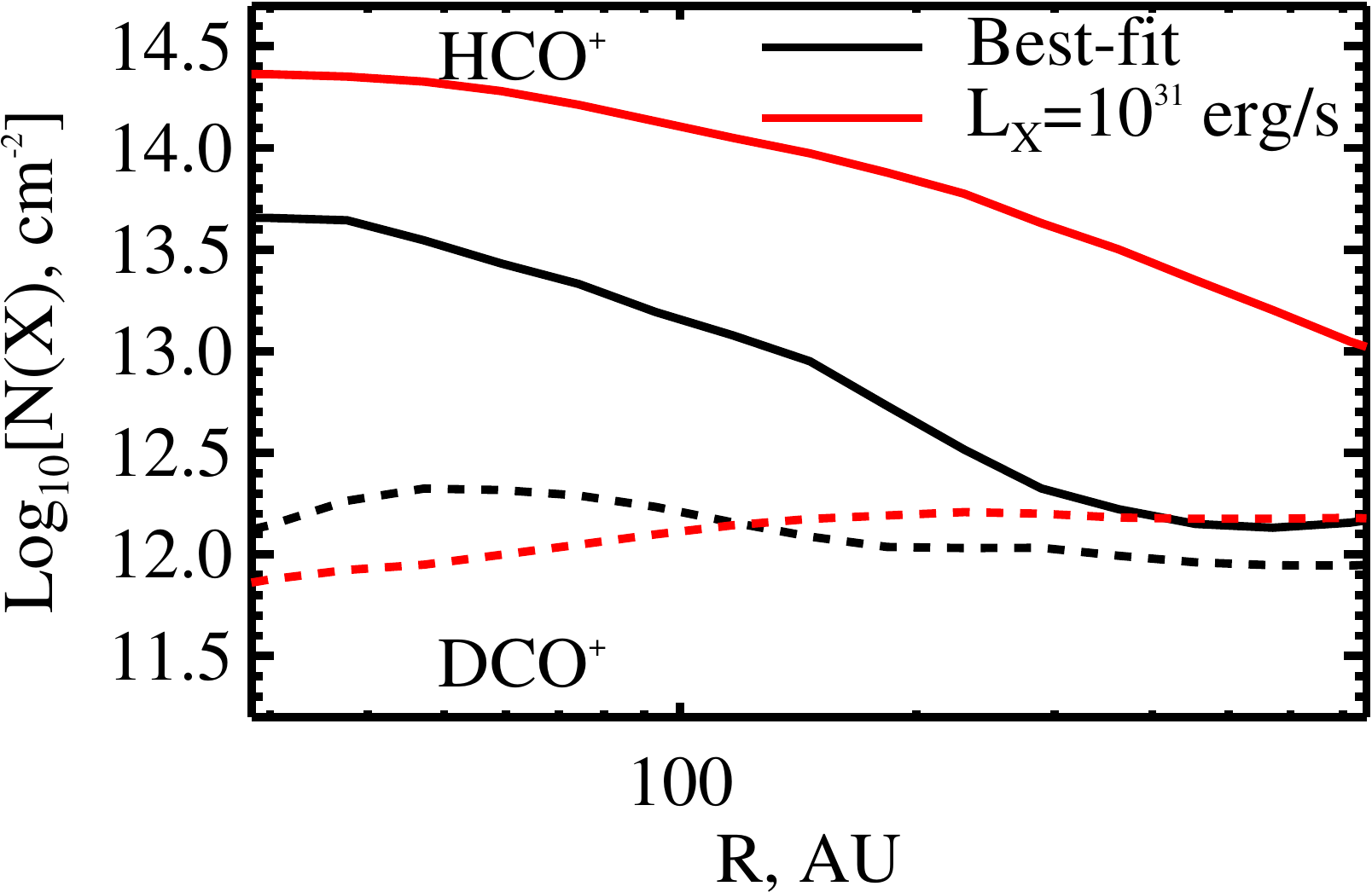}
   \caption{Increased X-Ray luminosity.}
   \label{fig:5highxrays}
   \end{subfigure}
   ~
   \begin{subfigure}[b]{0.32\textwidth}
   \includegraphics[width=\textwidth]{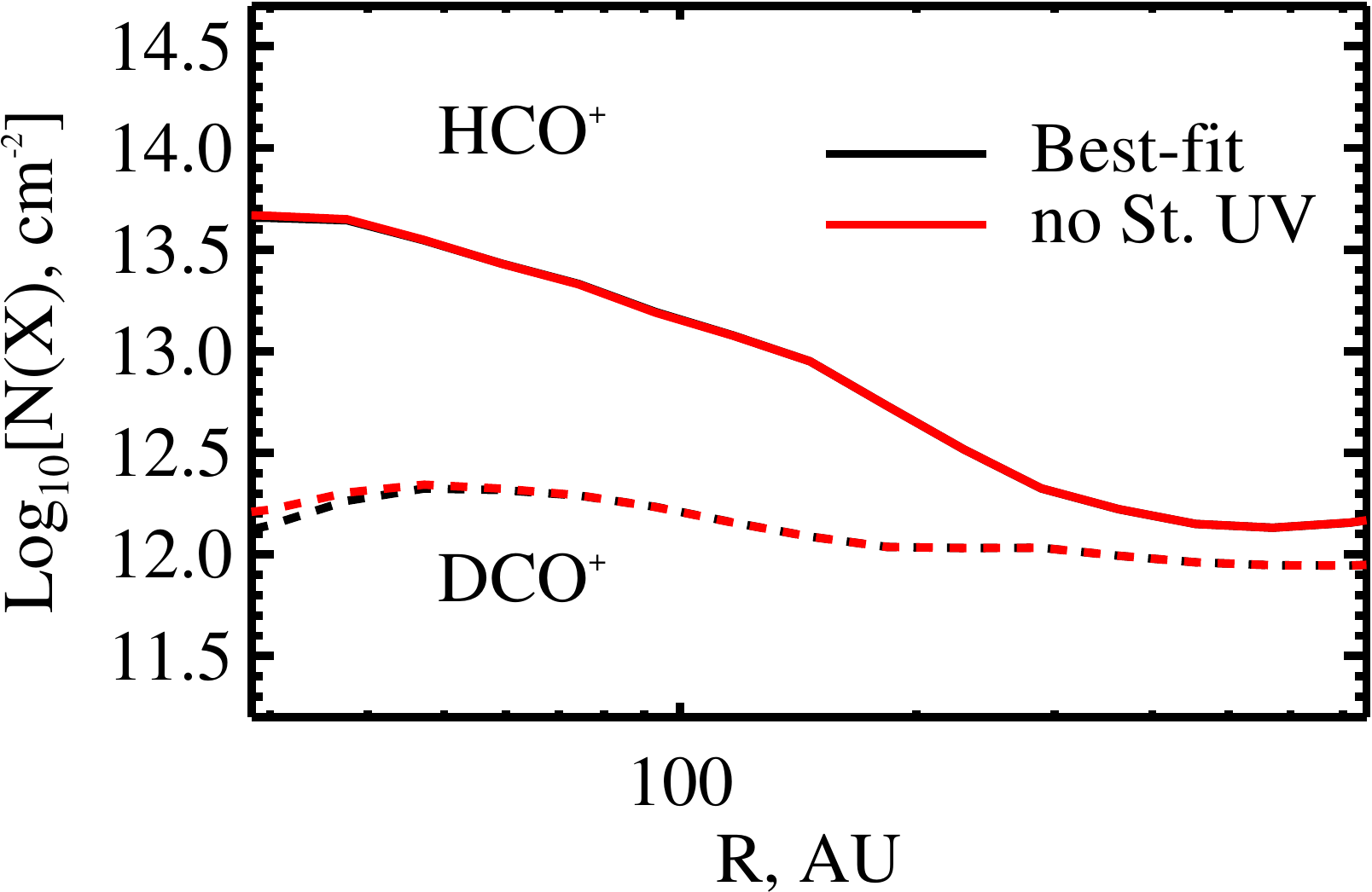}
   \caption{No stellar UV.}
   \label{fig:5stellar}
   \end{subfigure}
   
   ~\\
   
   \begin{subfigure}[b]{0.32\textwidth}
   \includegraphics[width=\textwidth]{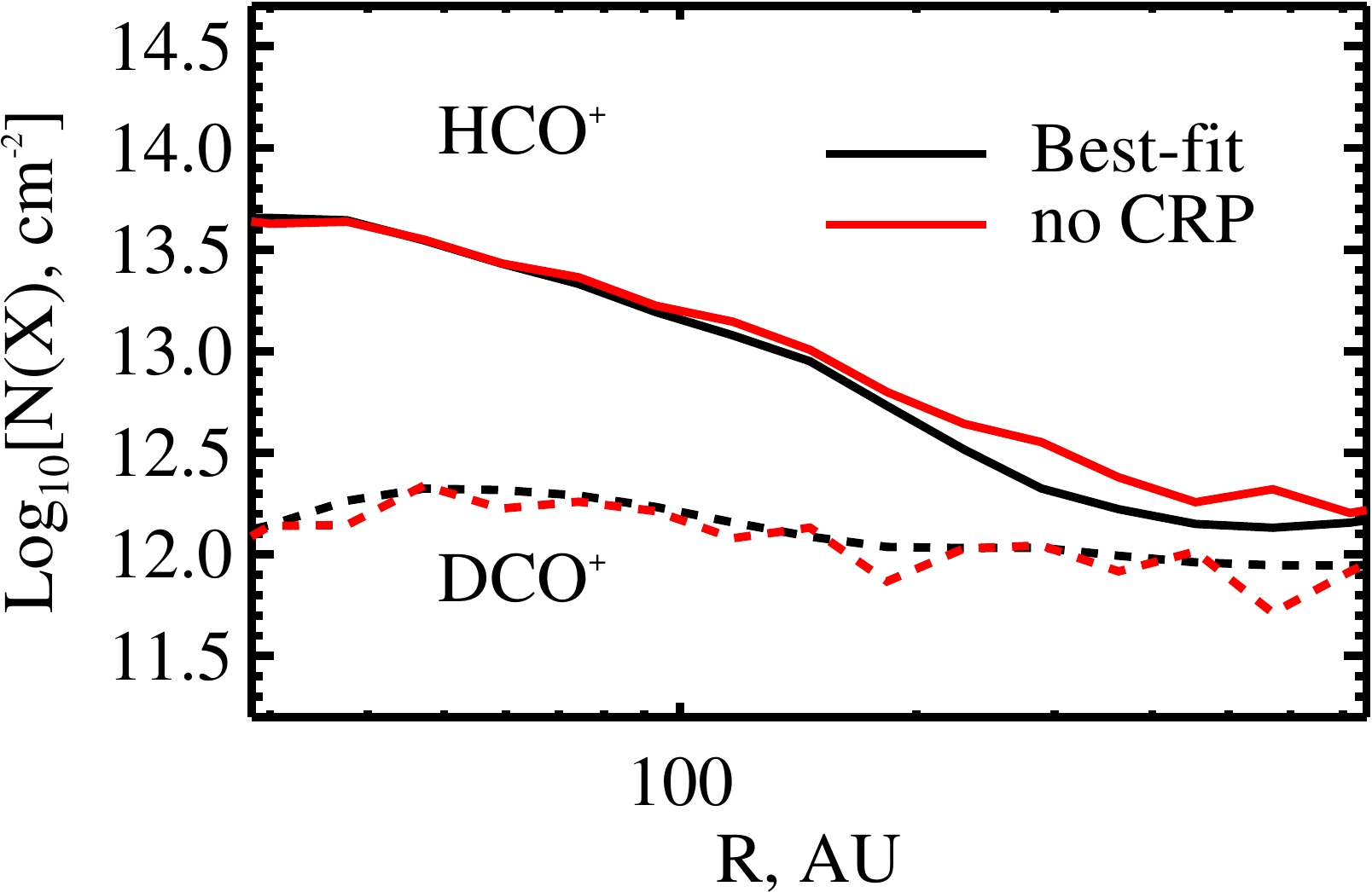}
   \caption{No CRPs.}\label{fig:5crps}
   \end{subfigure}
   ~
   \begin{subfigure}[b]{0.32\textwidth}
   \includegraphics[width=\textwidth]{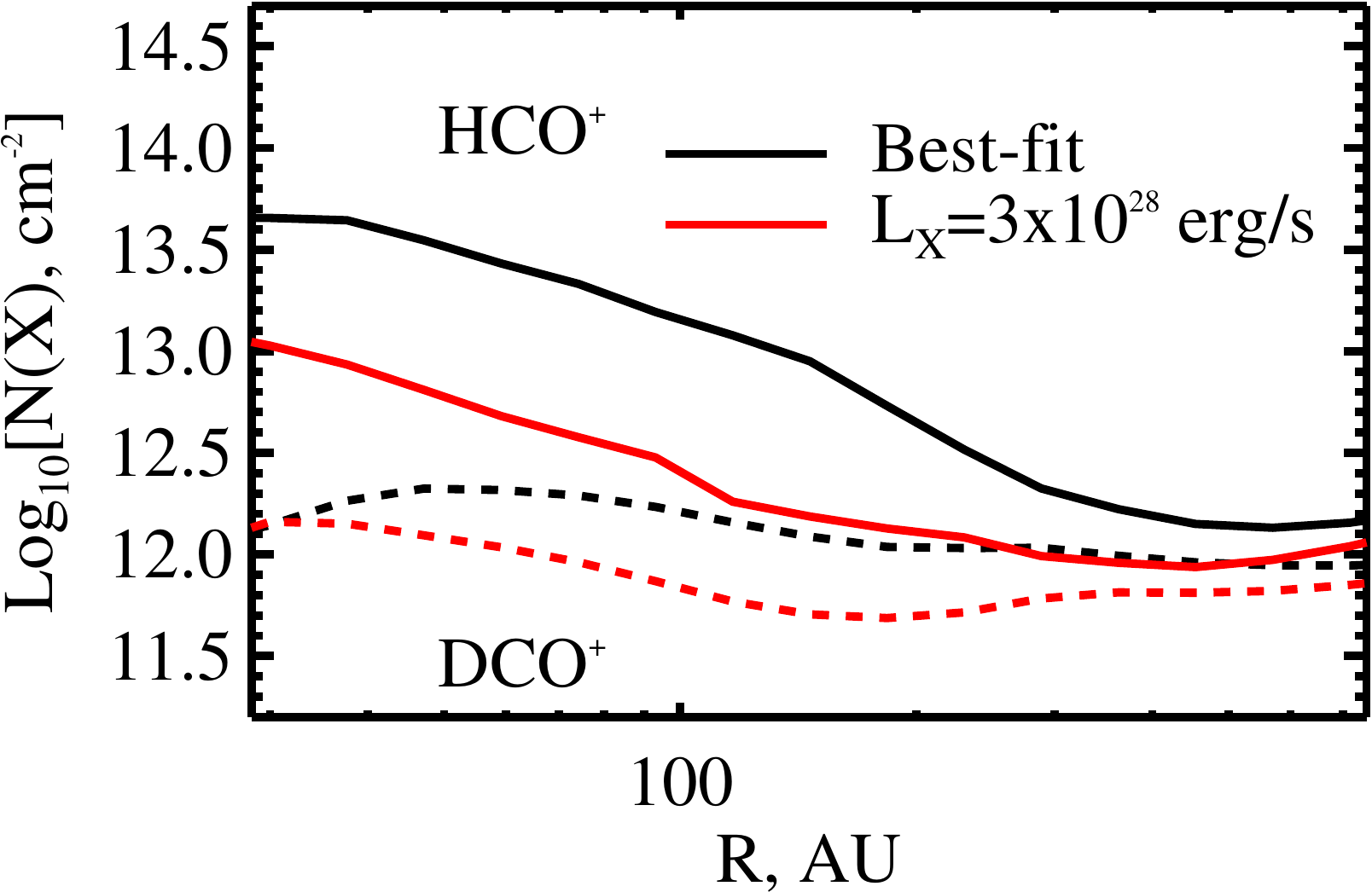}
   \caption{Reduced X-Ray Luminosity}
   \label{fig:5lowxrays}
   \end{subfigure}
   ~
   \begin{subfigure}[b]{0.32\textwidth}
   \includegraphics[width=\textwidth]{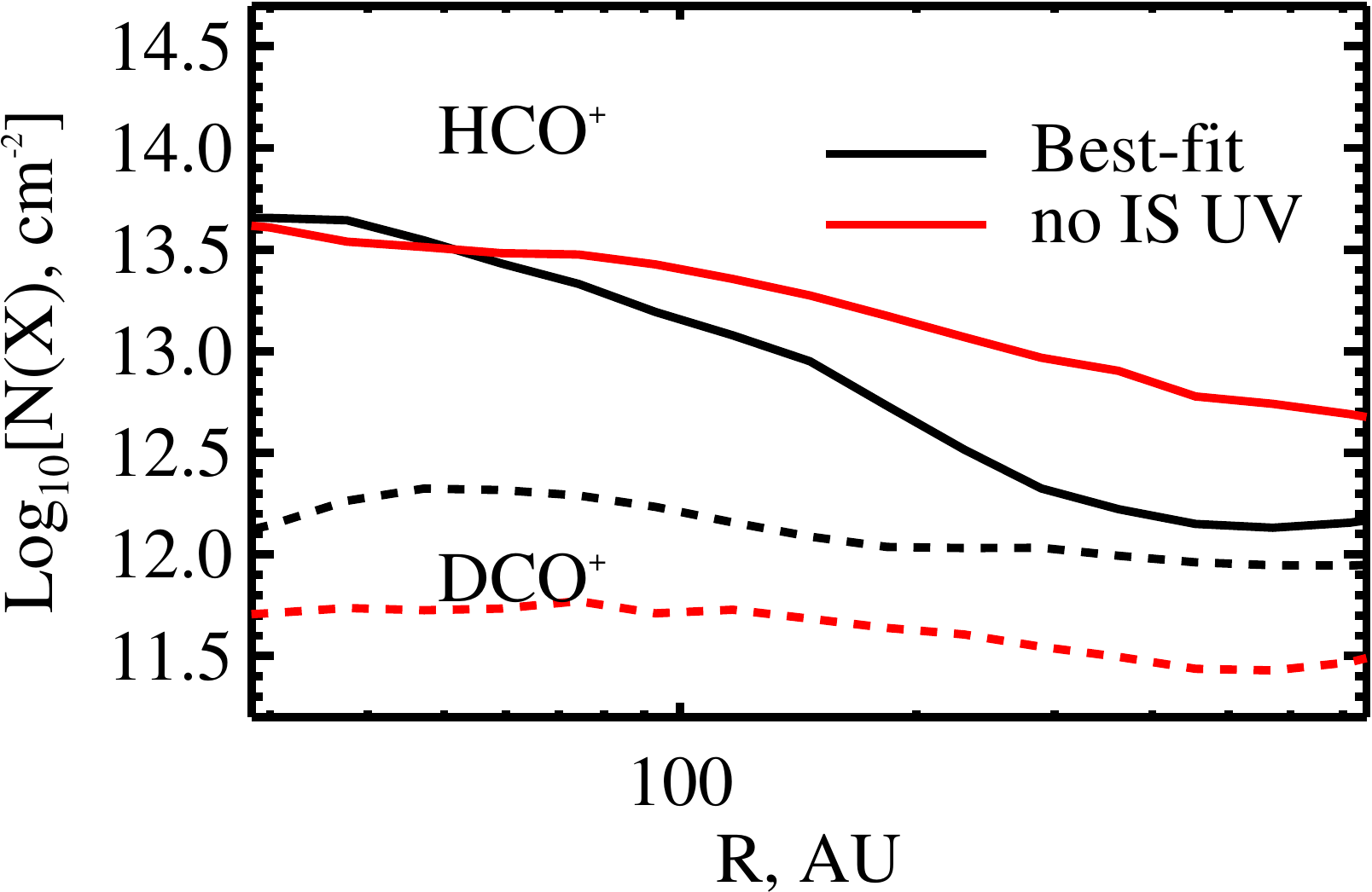}
   \caption{No interstellar UV.}
   \label{fig:5interstellar}
   \end{subfigure}
  
  \caption{Computed molecular column densities for DCO$^+$ (dashed line) and HCO$^+$ (solid line). 
Black lines show the columns obtained with the best-fit model (see Section~\ref{sec:bestfit_model}), 
red lines show the columns when the specified physical parameter is changed in the best fit model 
while the others are held constant. For reference, the canonical model has a single grain population of $a = 0.1~\mu \mathrm{m}$, $L_{\mathrm{X}} = 3 \times 10^{29}$~erg~s$^{-1}$, $\chi_*$(100~AU) = 410 and a CRP ionization rate of $\zeta = 1.3 \times 10^{-17}$~s$^{-1}$.}\label{fig:comp_abund}
\end{figure*}

Deuterium fractionation is typically used as a probe of the thermal history of an environment. In 
this section we explore physical parameters other than temperature which can alter the observed 
$R_{\rm D} \equiv N({\rm DCO}^+) /\, N({\rm HCO}^+)$ ratios in disks through a suite of chemical 
models derived from our DM~Tau model.

Figure \ref{fig:deuterium_frac_time} shows the radial profile of $R_{\rm D}$ found for DM~Tau 
observationally (dashed black line) and at various time steps in the model, $t = 
\{10^{3},\,10^{4},\,10^{5},\,10^{6},\, 5\times10^{6}\}$~yrs (solid blue lines). Also marked are 
values for TW~Hya (orange dash-dotted line; \citet{2008ApJ...681.1396Q}), HD~163296 (red dotted 
line; \citet{2013A&A...557A.132M}) and an average of prestellar cores (yellow hatched region; 
\citet{1995ApJ...448..207B,2002ApJ...565..344C}). Enhanced values relative to the cosmic abundance 
of [D]\,/\,[H] $\sim 10^{-5}$ are indicative of continued gaseous processing in a cold, $T \la 
20-30$~K, environment; a none too surprising conclusion given the kinetic temperature probed by the 
line emission, $T\sim 10-20$~K (see Table~\ref{tab:bestfit_res}). Furthermore, the radial increase 
is to be expected due to the radial temperature gradient in a disk. That is, the pace of deuterium 
fractionation and synthesis of the H$_3^+$ isotopologues hastens in colder outer disk regions. In 
turn, higher abundances of the H$_3^+$ isotopologues imply more efficient formation of DCO$^+$ in 
ion-molecule reactions with gaseous CO, increasing $R_{\rm D}$.

Our $R_{\rm D}$ values, ranging from $\approx 0.1 - 0.2$ between 50~AU and 430~AU, are almost two 
orders of magnitude higher than the disk average value of $(4.0 \pm 0.9) \times 10^{-3}$ found by 
\citet{2006A&A...448L...5G}, who used HCO$^+$ J = (1-0) interferomic data and DCO$^+$ J = (3-2) 
single dish data. This discrepancy can be explained by the assumptions made in the calculation of 
HCO$^+$ and DCO$^+$ column densities. Firstly, it was assumed that DCO$^+$ was radially co-spatial 
with the HCO$^+$ emission which extends out to $\sim 800$\,au, an outer radii similar to that of 
CO. This increased emitting region would result in a lower, disk average value for DCO$^+$, reducing 
$R_{\rm D}$. Secondly, we have shown that HCO$^+$ likely exhibits a complex molecular 
distribution including an inner hole in emission, see Sect. \ref{sec:observational_results}. When 
fitting a single transition this complexity was found to drive the parameterisation to favour 
steeper power laws describing column densities, consistent the HCO$^+$ J = (1-0) data from 
\citet{Pietu_ea07}, yielding a much smaller $R_{\rm D}$ value in the inner disk than found with our 
data.

In TW~Hydrae, another well studied protoplanetary disk, the value of $R_{\rm D}$ has been 
measured both as a disk average ($0.035 \pm 0.015$, \citet{2003A&A...400L...1V}), and with 
spatially resolved interferomic observations yielding a value from 0.01 to 0.1 between 30 and 90 au 
\citep{2008ApJ...681.1396Q}. The increase found in the outer regions of the disk when moving to 
spatially resolved data is consistent with our findings in DM Tau. Similarly, while the TW Hya disk 
is smaller and less massive than DM Tau, it appears to hold comparable values for $R_{\rm D}$ 
at similar distances to the star, namely around $\sim$ 0.1 at 100~AU.  More recently, 
\citet{2013A&A...557A.132M} used ALMA science verification data of HD 163296, an A1 spectral type 
Herbig Ae star and comprehensive modeling to ascertain local values of $R_{\rm D} \sim 0.3$ and a 
disk average of 0.02. Typically disks around Herbig Ae stars are warmer and more massive than those 
around classical T-Tauri stars, thus a reduced disk average value of $R_{\rm D}$ is to be expected.

Figure \ref{fig:deuterium_frac_time} also clearly shows continued enhancement of $R_{\rm D}$ up to $\sim$~1~Myr, indicative of continued processing of the gaseous CO during a disk lifetime. Thus, it is of no surprise that the observed value of 
$R_{\rm D}$ in DM~Tau, thought to be between 3 - 7~Myr, is higher than that found in low-mass 
prestellar cores ($0.045 \pm 0.015$, \citet{1995ApJ...448..207B}). 

Molecular layers in a protoplanetary disk are far beyond closed systems. Measurements of 
$R_{\rm D}$ through $N$(DCO$^+$)/\,$N$(HCO$^+$) are sensitive to more than just the deuterium 
fractionation efficiency, but are compounded by physical parameters which can alter the abundances 
of HCO$^+$ and DCO$^+$. In the remainder of this section we explore how else the physical 
environment can change the observed value of $R_{\rm D}$ by running a suite of chemical models and 
varying a single parameter. This will aid in analysis when comparing values of $R_{\rm D}$ from 
different astrophysical environments. To motivate the choice of parameters studied, we discuss 
briefly the process of deuterium fractionation.

Deuterium fractionation occurs due to the energy difference between deuterium and hydrogen atoms, 
thereby resulting in the deuteration of H$_3^+$ by HD to be exothermic:

\begin{equation}
 {\rm HD} + {\rm H}_3^+ \leftrightarrow {\rm H}_2 + {\rm H}_2{\rm D}^+ + \Delta E,
 \label{eq:deuteration}
\end{equation}

\noindent where $\Delta E = 232$~K \citep{2000A&A...361..388R}. Hence, low energy environments are 
conducive to enhanced abundaces of deuterated isotopologues resulting from successive deuteration 
of H$_3^+$. For H$_2$, $\Delta E$ corresponds to $T_{\rm kin} \approx 30$~K resulting in 
inefficient fractionation above this temperature. This is clearly evident in both observational and 
modelled results (see Fig.~\ref{fig:deuterium_frac_time}) which show an increase in $R_{\rm D}$ at 
larger radii where the disk is cooler. Furthermore, the larger internal energy of ortho-H$_2$ allows 
for greater leverage in the back reaction of Eqn.~\ref{eq:deuteration}. It has been shown that ortho 
/ para ratios $\gtrsim 0.1$ can `poison' the overall fractionation efficiency, even when the 
kinetic temperatures are low. Proton exchange due to collisions will reduce the initial 
H$_2$ ortho / para fraction of $\sim 0.75$. A reduced ortho-para fraction will increase the 
efficiency of deuterium fractionation, see Fig~\ref{fig:co_ice}a.
\citep[e.g.,][]{1992A&A...258..479P,2009A&A...494..623P,2011ApJ...729...15C,2013A&A...551A..38P,
2014ApJ...787...44A}. However, the relatively high densities of the molecular layer in a disk 
ensure that this is a relative quick process taking $\sim 10^{5}$~yrs.

\subsubsection{Regulating CO Abundance}

As a parental molecule of HCO$^+$ and DCO$^+$, the CO abundance is intimately linked to the 
abundances of HCO$^+$ and DCO$^+$. Gaseous CO must be sufficiently abundant to 
efficiently convert the H$_3^+$ and H$_2$D$^+$ into HCO$^+$ and DCO$^+$ respectively. However, CO 
readily freezes out at $T \approx 21$~K, vastly reducing the available reaction partners, see 
Fig~\ref{fig:co_ice}b. Therefore, DCO$^+$ is most efficiently produced where temperatures are high 
enough to maintain a relatively low level of CO depletion, yet cool enough to allow efficient 
fractionation. This is visible in Fig.~\ref{fig:deuterium_frac}a where $N$(DCO$^+$) peaks around 
50~AU. 

For the same mass of dust, larger grains have a reduced surface area onto which CO can freeze out, 
reducing the depletion of CO and hence expanding its molecular layer towards the disk midplane. 
Fig.~\ref{fig:5grains} shows the change in $N$(HCO$^+$) and $N$(DCO$^+$) when the grain size in our 
best fit model is increased to 1~$\mu$m. Outside the CO snowline $N$(HCO$^+$) is enhanced due to the 
greater availability of CO to react with H$_3^+$ resulting in a shallower gradient. On the other 
hand, $N$(DCO$^+$) is uniformly decreased. The decrease of DCO$^+$ abundances is associated with 
less efficient fractionation of H$_3^+$ in its molecular layer. Rapid ion-molecule reactions of 
H$_3^+$ with volatile molecules such as CO, which are less depleted from the gas phase, compete 
with fractionation processes and hence lower production of deuterated H$_3^+$ isotopologues. This 
results in a reduced $R_{\rm D}$ as a disk average and a weaker radial gradient. 

\subsubsection{Ionization}

The other parental molecules of HCO$^+$ and DCO$^+$ are H$_3^+$ and H$_2$D$^+$ respectively, both 
of which require the ionization of H$_2$. Figs~\ref{fig:5highxrays}-\subref{fig:5interstellar} show 
how the column densities are affected when ionization sources in the best fit model are altered.

Clearly shown in Figs.~\ref{fig:5stellar} and \ref{fig:5crps}, stellar UV and CRPs play 
little role in the abundance of these two species. In our disk model, UV scattering is neglected 
therefore the stellar UV radiation becomes quickly absorbed in radial direction by the dust, thus 
making little impact on the abundance of HCO$^+$ and DCO$^+$. Additionally, high energy CRPs can penetrate to the molecular layers, however they have such a low flux that they are of little consequence in the life-cycle 
of HCO$^+$ and DCO$^+$.

Conversely, the abundances of HCO$^+$ and DCO$^+$ are sensitive to the stellar X-rays, 
the dominant ionization source in the molecular layer. Increased values of $L_{X}$ lead to an 
enhanced abundance of HCO$^+$ across the entire disk, as shown by Fig.~\ref{fig:5highxrays}. 
DCO$^+$ production is suppressed in the inner disk, $r \lesssim 100$~AU, while outer regions display 
an enhancement. The DCO$^+$ production is suppressed in the inner disk despite an increase in 
overall H$_3^+$ isotopologue abundances within the inner molecular layer due to an 
increased ortho/para-H$_2$ fraction arising from the increased X-ray luminosity injecting sufficient energy into the disk for re-equilibriation through ion-molecule and nuclear spin-state processes. The increased 
ortho-fraction of the H$_3^+$ isotopologues slows the overall pace of the deuterium 
fractionation. In addition, an increase of DCO$^+$ abundances in the outer disk is smaller 
than for HCO$^+$ because the DCO$^+$ layer is located deeper into the disk, where temperatures 
favor deuterium enrichment andare better shielded from impinging stellar X-ray photons than 
the more extended HCO$^+$ molecular layer.

Reducing the stellar X-ray luminosity reduced both the HCO$^+$ and DCO$^+$ abundances. This is a 
more pronounced effect in the inner disk due to the oblique angles of incident X-rays, as found in 
\citet{Henning_ea10}. The higher sensitivity of HCO$^+$ to changing X-ray luminosities, again due 
to the difference in vertical extents of the molecular layers, is reflected in the gradient of 
$R_{\rm D}$; a lower $L_{X}$ leads to a less radially dependent $R_{\rm D}$. Additionally, HCO$^+$ 
production is suppressed to such a low level that $R_{\rm D}$ can reach $\sim 1$ in the outer disk. 

A more puzzling result is the strong influence of interstellar UV radiation on $R_{\rm D}$. In the 
disk model without an IS UV field, $R_{\rm D}$ becomes lower by up to an order of magnitude 
compared to the best-fit model. This is mainly due to a uniform decrease of the DCO$^+$ column 
density throughout the disk by a factor of $\sim 3$, and an increase of the HCO$^+$ column density 
at $r \ga 50-60$~AU by a similar factor of about 3, producing a near constant $R_{\rm D}$ across 
the radius of the disk.

The IS UV photons play two main roles for disk chemical evolution. Firstly, they partly contribute 
to the ionization and dissociation of disk matter in the molecular layer, particularly beyond $r\ga 
100$~AU. Secondly, and more importantly for $R_{\rm D}$, they bring heavy ices back to the gas phase 
by photodesorption and thus partly regulate surface processes. We found that in the DM~Tau model 
without IS UV radiation, CO gets more easily converted into CO$_2$ ice in the molecular layer at $T 
\la 30-40$~K through the slightly endothermic reaction of CO + OH + 80~K $\rightarrow$ CO$_2$ + H. CO$_2$ is unable to be photodesorbed and therefore partly dissociated in the gas. In addition, 
abundances of water ices increase, whereas abundances of atomic and molecular oxygen decrease. 
This leads to a drop in gas-phase CO abundances {\em locally} by a factor of 2-4 at all disk 
radii.

Furthermore, in the absence of photodesorption due to the absence of UV ionization, ions of alkali metals such as Na$^+$ and Mg$^+$, 
along with atomic ions such as S$^+$ and C$^+$, become less abundant and do not contribute 
considerably to the fractional ionization of the entire molecular layer. As a result, polyatomic 
ions like HCO$^+$ and H$_3$O$^+$ dominate the ionization structure, resulting in a decrease by a 
factor of a few in ionziation fraction due to their more efficient recombination with electrons. 
These two factors, lower ionization degree and lower CO abundances, lead to both lower DCO$^+$ 
abundances in the molecular layer and also reduced DCO$^+$ column densities.

Contrary to DCO$^+$, HCO$^+$ abundances and column densities show an increase at $r\ga 50$~AU 
in the DM~Tau model without IS UV. This is related to the fact that the DCO$^+$ molecular layer is 
narrower and located more deeply in the disk compared to the HCO$^+$ molecular layer. The lack of 
photodesorption and photodissociation in the upper part of the HCO$^+$ molecular layer increases 
abundances of CO and H$_3^+$ isotopologues, hence boosting production of HCO$^+$. This 
compensates for the decrease of gaseous CO due to its surface conversion into the CO$_2$ 
ice in the lower part of the HCO$^+$ molecular layer.

Comparisons with previously published models highlight the importance of considering these 
additional processes which alter the abundance of HCO$^+$ and DCO$^+$. Despite the relatively well 
understood deuterium fractionation mechanism there are orders-of-magnitude disparity between models. 
\citet{Aea02} modelled a smaller disk with gas extending on to 373~AU which exhibited a radial 
dependence of $R_{\rm D}$ varying between 0.003 to 0.06, values more more in accord with the smaller 
TW~Hya disk. Whereas a newer model of \citet{Willacy_07} found considerably higher values ranging 
from $R_{\rm D} = 0.1$ at 50~AU to reaching unity outside 100~AU, suggesting high $R_{\rm D}$ 
values are to be expected. 

Hence, while $N$(DCO$^+$)/\,$N$(HCO$^+$) provides an easily accessible measure of deuterium 
fractionation, a link through several environments in the cycle of molecular gas, other parameters, 
particularly the X-ray luminosity of the central star, the interstellar UV field and the grain 
evolution, are folded into this measurement. In the case of protoplanetary disks, \emph{the X-ray 
luminosity of the host star must be well constrained in order to fully characterise the deuterium 
fractionation present in the disk}.

%== Ionization ======================================================================
\subsection{Ionization Fraction, $x(e^-)$} 

% Ionization fraction in DM Tau.
\begin{figure}
 \centering
 \includegraphics[width=\columnwidth]{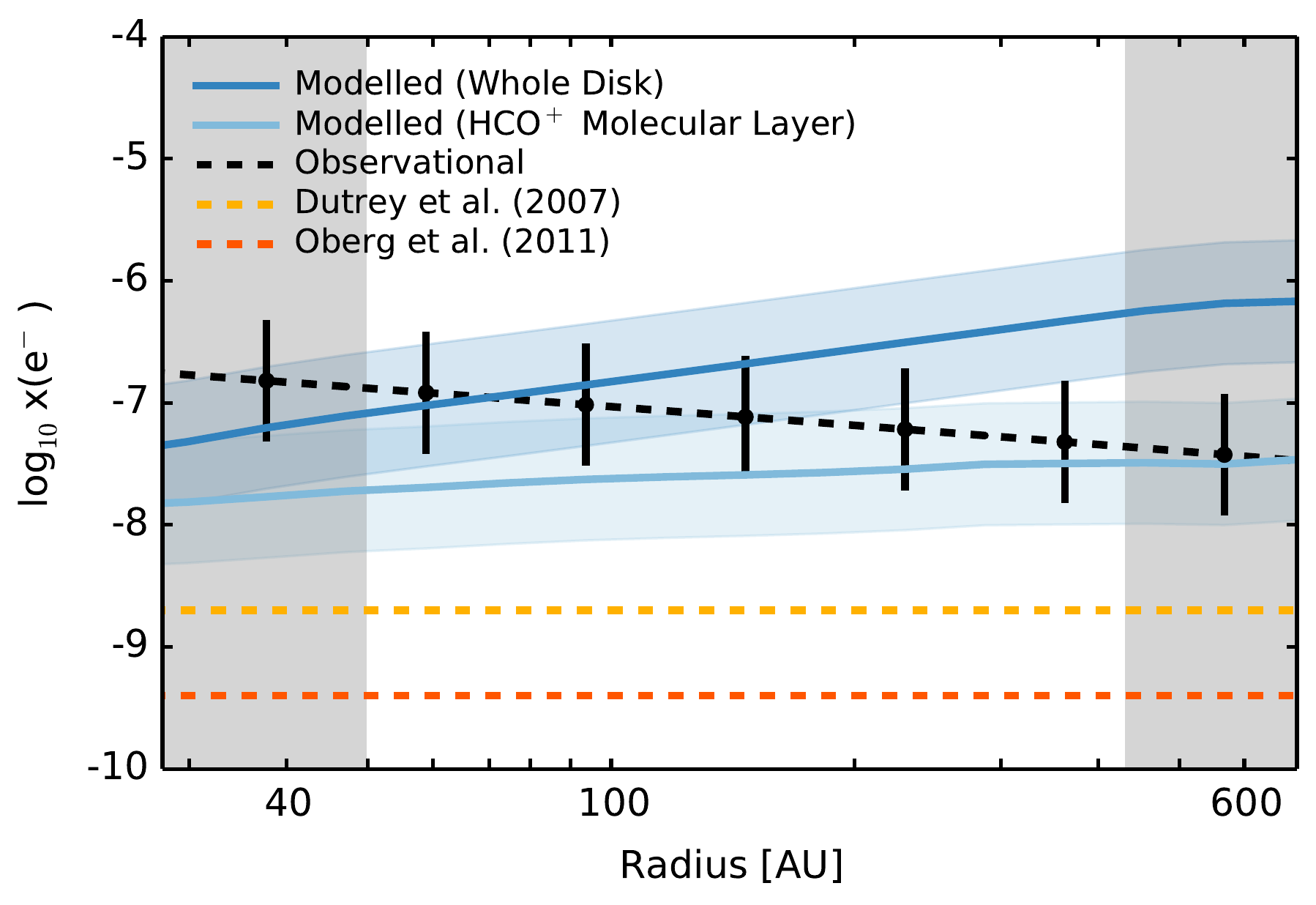}
 \caption{Constraints on the ionization fraction in DM~Tau. Results from 
the steady state approximation are shown with: observational values, black 
dashed; modelled values from the entire disk, dark blue solid; modelled 
values from the HCO$^+$ molecular layer, light blue solid; lower limit 
from \citet{2007A&A...464..615D}, light orange solid and lower limit from 
\citet{2011ApJ...743..152O}, dark orange solid. All errors, dominated by the values from the chemical modelling, are a factor of 3.}\label{fig:ion_fraction}
\end{figure}

% Ionization structure and dominant charge carriers.
\begin{figure*}
 \centering
 \includegraphics[width=1.9\columnwidth]{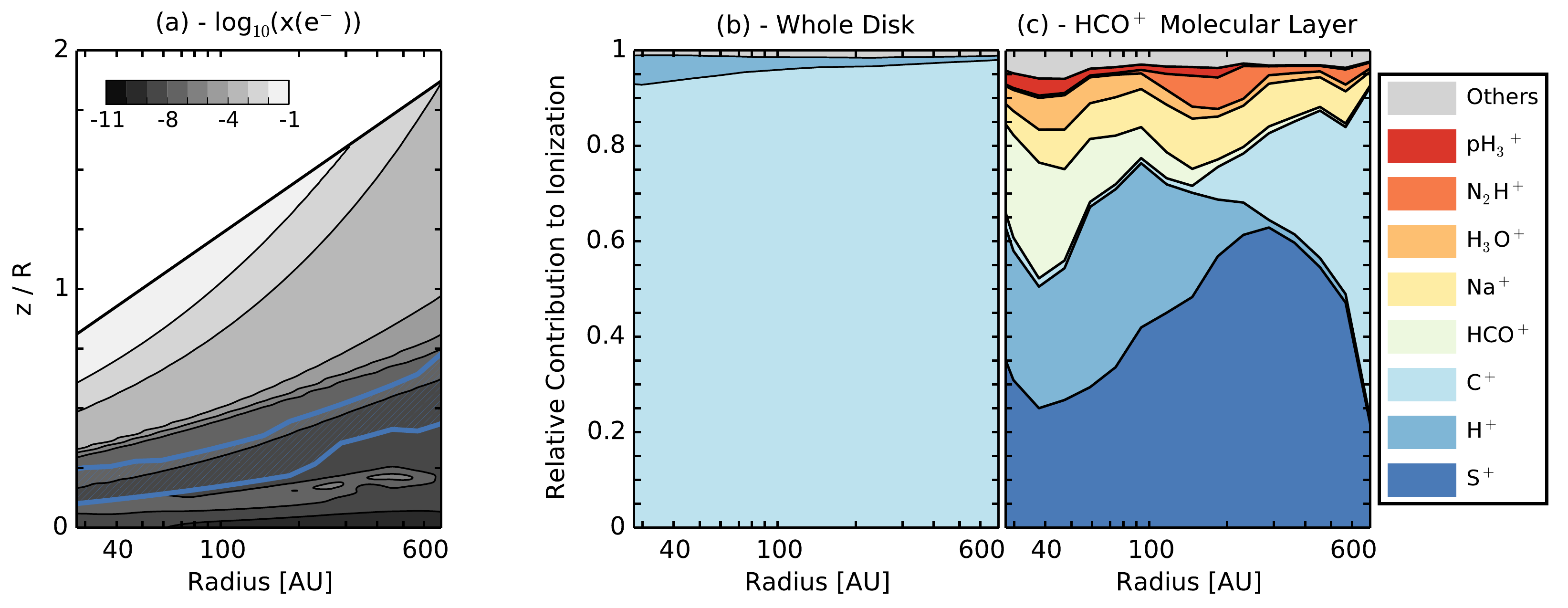}
 \caption{(a) Ionization structure of the DM~Tau disk model. The region bounded by blue lines is what we 
have defined as the HCO$^+$ molecular layer. (b) and (c) Relative contributions of charged species 
to the ionization level over the whole disk and the HCO$^+$ molecular layer. As a disk average, 
panel (b), it is clear the charge is dominated by C$^+$ and H$^+$, the gray region representing all 
other charged species. In the HCO$^+$ molecular layer, panel (c), defined as the region that 
contributes 90\% of the total HCO$^+$ column density, HCO$^+$ supplies a majority of the charge with 
large contributions from H$^+$, C$^+$ and S$^+$.}\label{fig:ion_abundance}
\end{figure*}

% Reactions involved with ionization. 
\begin{table}
\centering                         
\footnotesize
\onehalfspacing
\begin{tabular}{ccccc}        
\hline\hline
\textsc{Reaction} & \textsc{Rates} & $\alpha$ \textsc{(cm$^3$\,s$^{-1}$)} & $\beta$ & $\gamma$ 
\textsc{(k)}\\ \hline
\multicolumn{5}{c}{\textsc{deuteration}}\\
H$_3^+$ + HD $\leftrightarrow$ H$_2$D$^+$ + H$_2$ & $k_1, k_{-1}$ & $1.7 \times 10^{-9}$ & 0 & 220 
\\
H$_2$D$^+$ + HD $\leftrightarrow$ D$_2$H$^+$ + H$_2$ & $k_2, k_{-2}$ & $8.1 \times 10^{-10}$ & 0 & 
187 \\
D$_2$H$^+$ + HD $\leftrightarrow$ D$_3^+$ + H$_2$ & $k_3, k_{-3}$ & $6.4 \times 10^{-10}$ & 0 & 234 
\\
\multicolumn{5}{c}{\textsc{ion--molecule}}\\
H$_3^+$ + CO $\rightarrow$ HCO$^+$ + H$_2$ & $k_{\mathrm{CO}}$ & $1.61 \times 10^{-9}$ & 0 & - \\
H$_2$D$^+$ + CO $\rightarrow$ HCO$^+$ + HD & $k_{\mathrm{CO}}$ & $1.61 \times 10^{-9}$ & 0 & - \\
H$_2$D$^+$ + CO $\rightarrow$ DCO$^+$ + H$_2$ & $k_{\mathrm{CO}}$ & $1.61 \times 10^{-9}$ & 0 & - \\
D$_2$H$^+$ + CO $\rightarrow$ HCO$^+$ + D$_2$ & $k_{\mathrm{CO}}$ & $1.61 \times 10^{-9}$ & 0 & - \\
D$_2$H$^+$ + CO $\rightarrow$ DCO$^+$ + HD & $k_{\mathrm{CO}}$ & $1.61 \times 10^{-9}$ & 0 & - \\
D$_3^+$ + CO $\rightarrow$ DCO$^+$ + D$_2$ & $k_{\mathrm{CO}}$ & $1.61 \times 10^{-9}$ & 0 & - \\
\multicolumn{5}{c}{\textsc{recombination}}\\
H$_3^+$ + e$^-$ $\rightarrow$ various & $k_{\mathrm{rec0}}$ & $6.8 \times 10^{-8}$ & -0.5 & - \\
H$_2$D$^+$ + e$^-$ $\rightarrow$ various & $k_{\mathrm{rec1}}$ & $6.0 \times 10^{-8}$ & -0.5 & - \\
D$_2$H$^+$ + e$^-$ $\rightarrow$ various & $k_{\mathrm{rec2}}$ & $6.0 \times 10^{-8}$ & -0.5 & - \\
D$_3^+$ + e$^-$ $\rightarrow$ various & $k_{\mathrm{rec0}}$ & $2.7 \times 10^{-8}$ & -0.5 & - \\
\hline                          
\end{tabular}
\singlespacing
\caption{Reactions involved in the considered steady state system and the rates used in our 
modeling. Forward rates should be read as $\alpha (T/300)^{\beta}$ and backwards as $\alpha 
(T/300)^{\beta}\exp(-\gamma/T)$. Adapted from \citet{2008A&A...492..703C}.}
\label{tab:ioniz_chemistry}
\end{table}

HCO$^+$ is often touted as the most dominant ion in the warm molecular layer of a protoplanetary 
disk. As such, it is frequently used as a proxy of the ionization in this region 
\citep{2004A&A...417...93S,2007A&A...464..615D,2008ApJ...681.1396Q,2011ApJ...743..152O}. However, 
the large radial and vertical gradients in physical parameters characteristic of a protoplanetary 
disk introduce several complications in deriving knowledge of the ionization fraction from a single 
charged species. Common practice therefore is to make a steady state approximation, a methodology 
that has been applied to a range of astrophysical scales: from protoplanetary disks to molecular 
clouds and supernova remnants 
\citep{1977ApJ...217L.165G,2002P&SS...50.1133C,2008A&A...492..703C,2014A&A...568A..50V}.

Introduced by \citet{1977ApJ...217L.165G}, this assumes a heavily reduced 
chemical network in a steady state of ionization. As discussed previously, the abundance of HCO$^+$ 
is largely governed by two main processes: creation through ion-neutral reactions between H$_3^+$ 
and CO and destruction via electronic recombination. Similar pathways hold for DCO$^+$ but with the 
deuterated H$_2$D$^+$ in place of H$_3^+$. \citet{2002P&SS...50.1133C} showed the chemical kinetics 
of such a network in steady state can be reduced to:

\begin{equation}
x(e^-) = \frac{1}{k_{{\rm rec}1}} \left( \frac{k_1 x({\rm HD})}{3 R_{\rm D}} - k_3 x({\rm CO})  
\right),
\end{equation}

\noindent where the rates and associated reactions are found in Table \ref{tab:ioniz_chemistry}. 

This requires knowledge of both $N$(HD), of which we only have one observation of in a 
protoplanetary disk, TW~Hya \citep{2013Natur.493..644B}, and the total gas column from which to 
convert from column density to relative abundances, a value that cannot be well constrained 
observationally without several assumptions \citep{2010A&A...518L.125T}. Application to DM~Tau is 
thus limited to the case where we must assume values of $x$(CO) and $x$(HD). Note that this has 
since been expanded to include all multiply-deuterated isotopologues of H$_3^+$ and charged grains, 
however the lack of observational constraints on these would further compound the issues detailed 
above in the case of a protoplanetary disk \citep{2008A&A...492..703C}. Note, however, that physical parameters derived from line emission will be of the molecular region and not applicable to the disk as a whole.

Using our observationally derived $R_{\rm D}$ values and values $x({\rm HD}) = 2.40 \times 10^{-5}$ and $x({\rm CO}) = 7.24 \times 10^{-5}$ taken from the best fit model, the later consistent with previous observations of DM~Tau \citep{Pietu_ea07}, we find an ionization fraction of $x(e^-) \sim 10^{-7}$, as shown by the dashed black line in 
Fig.~\ref{fig:ion_fraction}. This is consistent with the lower limits placed by 
\citet{2007A&A...464..615D} (light orange) and \citet{2011ApJ...743..152O} (dark orange). Blue lines 
show $x(e^-)$ from our best fit model, the light blue considering molecular column densities integrated over the warm molecular layer probed by our HCO$^+$ and DCO$^+$ observations\footnote{We define the molecular layer of a molecule such that the column density of that molecule contained in it is equal to 90~\% of that molecule's total column density. It is centred at the position which has a largest fractional contribution to the total column density.}, and those integrated over the whole disk in dark blue. Both of which qualitatively agree with the steady state values.  \citet{2008ApJ...681.1396Q} found a similar ionization fraction of $x(e^-) \sim 10^{-7}$ in TW~Hya when using the same steady state approximation.

However, disk ionization, is controlled by a myriad of atomic and molecular ions as shown by 
Fig~\ref{fig:ion_abundance}. Panels (b) and (c) show the relative contribution of the 
top eight most abundant ions as a function of radius for the whole disk (b) and the HCO$^+$ molecular 
layer (c). It is clear that ionization as a whole is dominated by the atomic ions C$^+$ and H$^+$ 
which contribute $\gtrsim 99~\%$ of the charge. Even within the molecular layer, atomic ions are 
the dominant charge carries with S$^+$, H$^+$ and C$^+$ contributing between 50~\% and 90~\% of the 
total charge. While HCO$^+$ is the dominant molecular ion, it contributes at most 
$\sim~20~\%$ of the charge in the inner regions and is severely depleted in the outer disk, $R 
\gtrsim 200$~AU. Thus, while HCO$^+$ being the most dominant \emph{molecular} ion in the disk, even 
in the molecular layer its contribution to total charge is dwarfed by that of atomic ions such as  
C$^+$, H$^+$ and S$^+$.

This additional source of ionization not considered in the steady state approximation can 
contribute to the difference in observed values of $x(e^-)$. Furthermore, it is surprising that 
observations sensitive to only a small region in the disk are able to recover disk average values 
relatively well. This is due to the density gradient towards the midplane; the disk average will 
draw heavily from values closer to the midplane.

%== Summary ======================================================================

\section{Summary}
\label{sec:summary}
In this paper Plateau de Bure Interferometer observations of the abundant molecules HCO$^+$, J = 
(3-2), (1-0) and DCO$^+$, J = (3-2) of DM~Tau in parallel with a suit of thermo-chemical models 
provided the framework with which to study deuterium fractionation and the ionization fraction in 
DM~Tau.

Using combined $\chi^2$-minimization and MCMC fitting techniques, we fitted a parametric model to 
the observations. \emph{HCO$^+$ was found to exhibit a complex emission structure}: (3-2) emission 
had a peak intensity at $r \approx 50$~AU and was considerably less extended than the (1-0) 
emission. HCO$^+$ (1-0) emission was found to be co-spatial with CO emission \citep{Pietu_ea07}. An 
inner hole of $r \approx 50$~AU in HCO$^+$ is needed to recreate the observations. Simultaneously 
fitting the (3-2) and (1-0) lines required the assumption that both lines had the same excitation 
temperature. DCO$^+$ was also tentatively found to peak at $r \approx 70$~AU, consistent with the CO 
snowline, however higher resolution observations are required to confirm this.

We find \emph{$R_{\rm D} = N({\rm HCO}^+)/N({\rm DCO}+)$ varies from 0.1 - 0.2 between 50 and 
430~AU}, values considerably higher 
than both the cosmic abundance $\sim$~10$^{-5}$ and those found in prestellar cores $0.035 \pm 
0.015$.  This is indicative of continued fractionation throughout the disk lifetime. Both TW~Hya and 
HD~163296 exhibit similar high levels of deuteration, the later peaking at $R_{\rm D} \sim 0.3$ 
\citep{2008ApJ...681.1396Q,2013A&A...557A.132M}. 

Through chemical modeling we have also explored other physical parameters which can affect $R_{\rm 
D}$. \emph{The most influential parameters are the level of 
interstellar UV and the X-ray luminosity 
of the central star}; the dominant ionization sources in the molecular layer. X-rays interact with 
$R_{\rm D}$ in a relatively straight forward manner: higher luminosities mean increased abundances 
of HCO$^+$ and DCO$^+$ and conversely for reduced luminosities. On the other hand, interstellar UV 
is more complex as it does not directly affect HCO$^+$ and DCO$^+$ but rather CO abundances. It 
enhances the DCO$^+$ abundance across the entire disk, while suppresses the formation of HCO$^+$ in 
the outer, $r \gtrsim 50$~AU, disk. These effects can be disentangled through the dependence of 
$R_{\rm D}$ on radius with the later producing a more radially constant value of $R_{\rm D}$.

Assuming a steady state system we estimate the electron fraction of the HCO$^+$ molecular layer to 
be $x(e^-) \sim 10^{-7}$, consistent with lower limits from \citet{2007A&A...464..615D} and
\citet{2011ApJ...743..152O}. This values is high enough to induce MRI turbulence. An analysis of the 
dominant charge carries in the molecular layer show that HCO$^+$ is the most dominant 
\emph{molecular} ion, however atomic ions are considerably more dominant in all regions of the 
disk. Thus constraints on ionization from the abundance of HCO$^+$ must take this into account.

%== Acknowledgments ==============================================================
\begin{acknowledgements}
R.T. is a member of the International Max Planck Research School for Astronomy and Cosmic Physics at 
the University of Heidelberg, IMPRS-HD, Germany. We thank the Plateau de Bure staff for performing 
the observations and helping with the data reduction. D. Semenov acknowledges support by the {\it 
Deutsche Forschungsgemeinschaft} through SPP~1385: ``The first ten million years of the solar 
system - a planetary materials approach'' (SE 1962/1-3).  This research made use of NASA's 
Astrophysics Data System. V.W. research is funded by the ERC starting grant 3DICE (grant agreement 336474).
\end{acknowledgements}

%== Bibliography ===================================================================
\bibliographystyle{aa}
\bibliography{bibliography}

\end{document}